\documentclass[twocolumn,showpacs,amsmath,amssymb,prl,footinbib,floatfix,superscriptaddress]{revtex4}

\usepackage{graphicx}% Include figure files
\usepackage{dcolumn}% Align table columns on decimal point
\usepackage{bm}% bold math
\usepackage{helvet}
\usepackage{amssymb}
\usepackage{amsmath}
\usepackage{amsfonts}
\usepackage{color}
\usepackage{verbatim}

\newcommand{\beq}{\begin{eqnarray}}
\newcommand{\eeq}{\end{eqnarray}}

\newcommand{\q}[1]{$Q_#1$}

\date{\today}
\begin{document}
%\title{Arrays of Superconducting Qubits for Dipolar Quantum Spin Models}
\title{Dipolar Spin Models with Arrays of Superconducting Qubits}

\author{M. Dalmonte}
\affiliation{Institute for Quantum Optics and Quantum Information of the Austrian Academy of Sciences,A-6020 Innsbruck, Austria}
\affiliation{Institute for Theoretical Physics, University of Innsbruck, A-6020 Innsbruck, Austria}
\author{S.I. Mirzai}
\affiliation{Institute for Quantum Optics and Quantum Information of the Austrian Academy of Sciences,A-6020 Innsbruck, Austria}
\affiliation{Institute for Experimental Physics, University of Innsbruck, A-6020 Innsbruck, Austria}
\author{P.R. Muppalla}
\affiliation{Institute for Quantum Optics and Quantum Information of the Austrian Academy of Sciences,A-6020 Innsbruck, Austria}
\affiliation{Institute for Experimental Physics, University of Innsbruck, A-6020 Innsbruck, Austria}
\author{D. Marcos}
\affiliation{Institute for Quantum Optics and Quantum Information of the Austrian Academy of Sciences,A-6020 Innsbruck, Austria}
\affiliation{Institute for Theoretical Physics, University of Innsbruck, A-6020 Innsbruck, Austria}
\author{P. Zoller}
\affiliation{Institute for Quantum Optics and Quantum Information of the Austrian Academy of Sciences,A-6020 Innsbruck, Austria}
\affiliation{Institute for Theoretical Physics, University of Innsbruck, A-6020 Innsbruck, Austria}
\author{G. Kirchmair}
\affiliation{Institute for Quantum Optics and Quantum Information of the Austrian Academy of Sciences,A-6020 Innsbruck, Austria}
\affiliation{Institute for Experimental Physics, University of Innsbruck, A-6020 Innsbruck, Austria}

%%%%%%
\begin{abstract}
We propose a novel platform for quantum many body simulations of dipolar spin models using current circuit QED technology. Our basic building blocks are 3D Transmon qubits where we use the naturally occurring dipolar interactions to realize interacting spin systems. This opens the way toward the realization of a broad class of tunable spin models in both two- and one-dimensional geometries. We illustrate the potential offered by these systems in the context of dimerized Majumdar-Ghosh-type phases, archetypical examples of quantum magnetism, showing how such phases are robust against disorder and decoherence, and could be observed within state-of-the-art experiments. 
\end{abstract}

%42.50.Dv Quantum state engineering and measurements
%75.10.Pq: Spin chain models
%85.25.-j 	Superconducting devices
%42.50.Pq 	Cavity quantum electrodynamics; micromasers
%03.67.Ac 	Quantum algorithms, protocols, and simulations

%%%%%%
\maketitle
%%%%%

\paragraph{Introduction. --}
In the present work we propose and analyze a novel setup for an analog quantum simulator of quantum magnetism using superconducting qubits. The scheme builds on the remarkable recent developments in Circuit QED~\cite{Dewes:2012ff,Reed:2012ud,vanLoo2013,Riste:2013,Roch:2014,Barends:2014} in the context of quantum simulation~\cite{Houck:2012,Viehmann2013a,Viehmann2013b,Mezzacapo:2014bs,Naether:2014dz}, and especially the 3D Transmon qubit~\cite{Paik:2011,Rigetti:2012}. The scheme promises a faithful implementation of many-body spin-$1/2$ Hamiltonians involving tens of qubits using state-of-the-art experimental techniques. The central idea behind the present work is to exploit the naturally occurring dipolar interactions between qubits to engineer the desired spin-spin interactions. In combination with the flexibility offered by solid-state setups for realizing arbitrary geometry arrangements, this allows us to design general dipolar spin models in ladder and 2D geometries. As we will show, our scheme competes favorably with present and envisaged quantum simulation setups for magnetism with cold atoms and trapped ions~\cite{Bloch:2012, Blatt:2012,Georgescu:2014kq}, and enables us to address some of the key challenges of quantum simulation including equilibrium and non-equilibrium (quench) dynamics~\cite{Polkovnikov:2011vn}. Moreover, we note that exploiting dipolar interactions to design dipolar spin models is conceptually different, and complementary to the remarkable recent experiments with superconducting circuits toward realizing the superfluid-Mott insulator transition, based on {\it wiring up} increasingly complex circuits of superconducting stripline cavities~\cite{Houck:2012}.

\begin{figure}[h]
\includegraphics[width = 0.96\columnwidth]{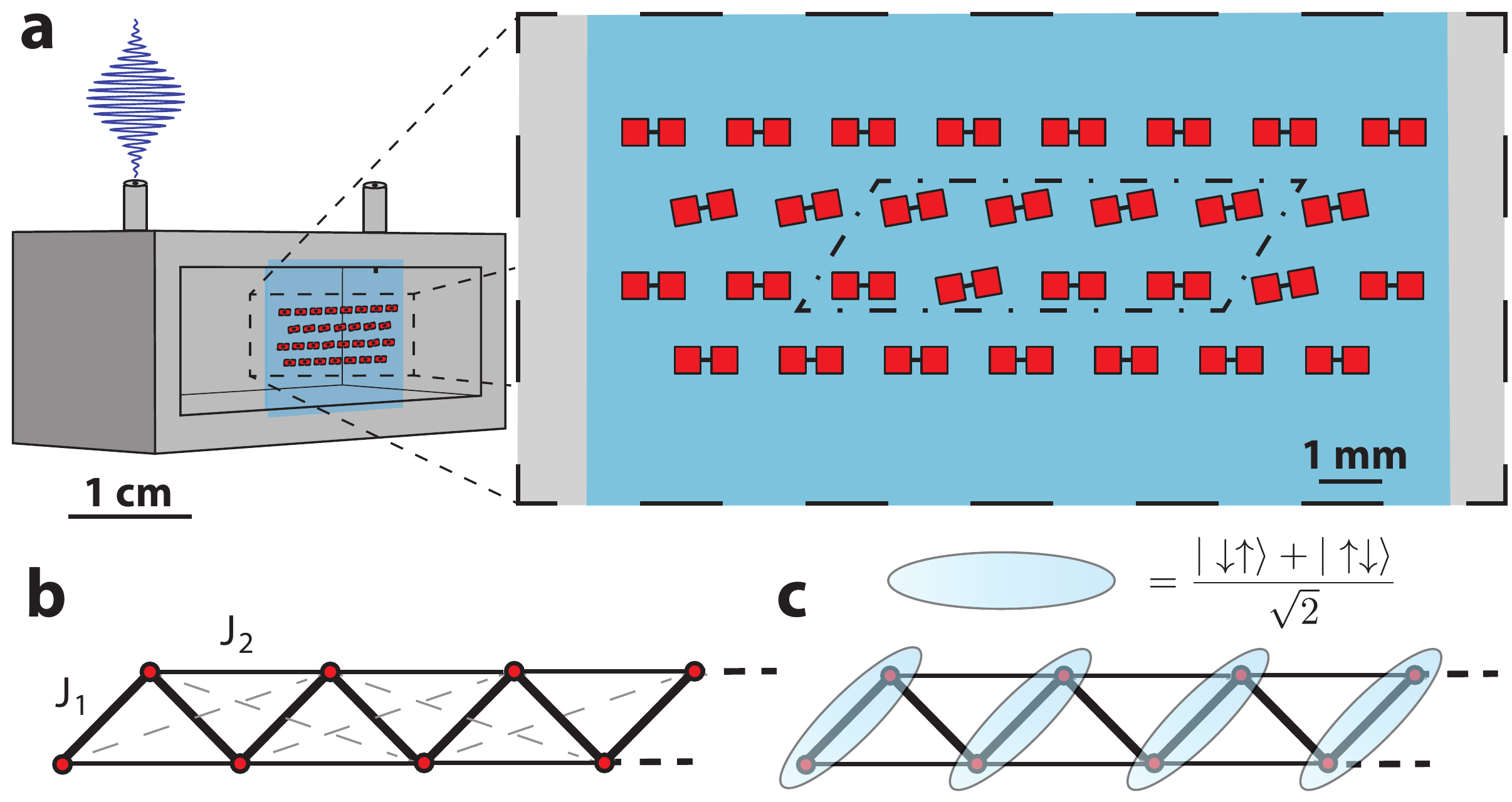}
\caption{Circuit QED implementation of spin models. (a) Drawing of one half of a waveguide cavity with two microwave couplers. Inside the cavity is a piece of sapphire with multiple Transmon qubits arranged on a triangular lattice. The rotation angles of the qubits are chosen such, that only a few of them couple to the fundamental mode of the cavity with a predetermined coupling strength. (b) The cavity can be loaded with an arbitrary subset of qubits e.g. considering the ones in the dot-dashed box in (a) realizes a triangular ladder. Here, the bond thickness denotes the interaction strength of nearest-neighbor ($J_1$) and next-nearest-neighbor ($J_2$) interactions. Additional longer-range contributions are represented by dashed lines (only the strongest are shown). (c) Sketch of the dimerized phase of the extended XY model in Eq.~\eqref{eqH}. The phase can be understood as a solid of local triplet states, denoted by the shaded areas. 
}
\label{fig_system} 
\end{figure}

In our analysis we address two of the key aspects of the design of our proposed simulator for quantum magnetism. First, we present a feasibility study of state-of-the-art experimental setups: this includes a discussion of the general mechanism to generate dipolar interactions between 3D Transmons, combined with {\it ab initio} simulations of the coupling strength in our spin model for various geometries. Second, we illustrate how state-of-the-art setups, composed of up to a dozen qubits and characterized by typical disorder and decoherence rates, are already able to demonstrate paradigmatic signatures of quantum magnetism. In particular, we show how a dimerized phase~\cite{lacroix2011introduction}, a valence-bond-solid reminiscent of the Majumdar-Ghosh state widely discussed in the context of quantum spin chains~\cite{Majumdar1969}, can be realized and probed with current technologies. 

\begin{figure}[t]
%\begin{center}
\includegraphics[width = 0.98\columnwidth]{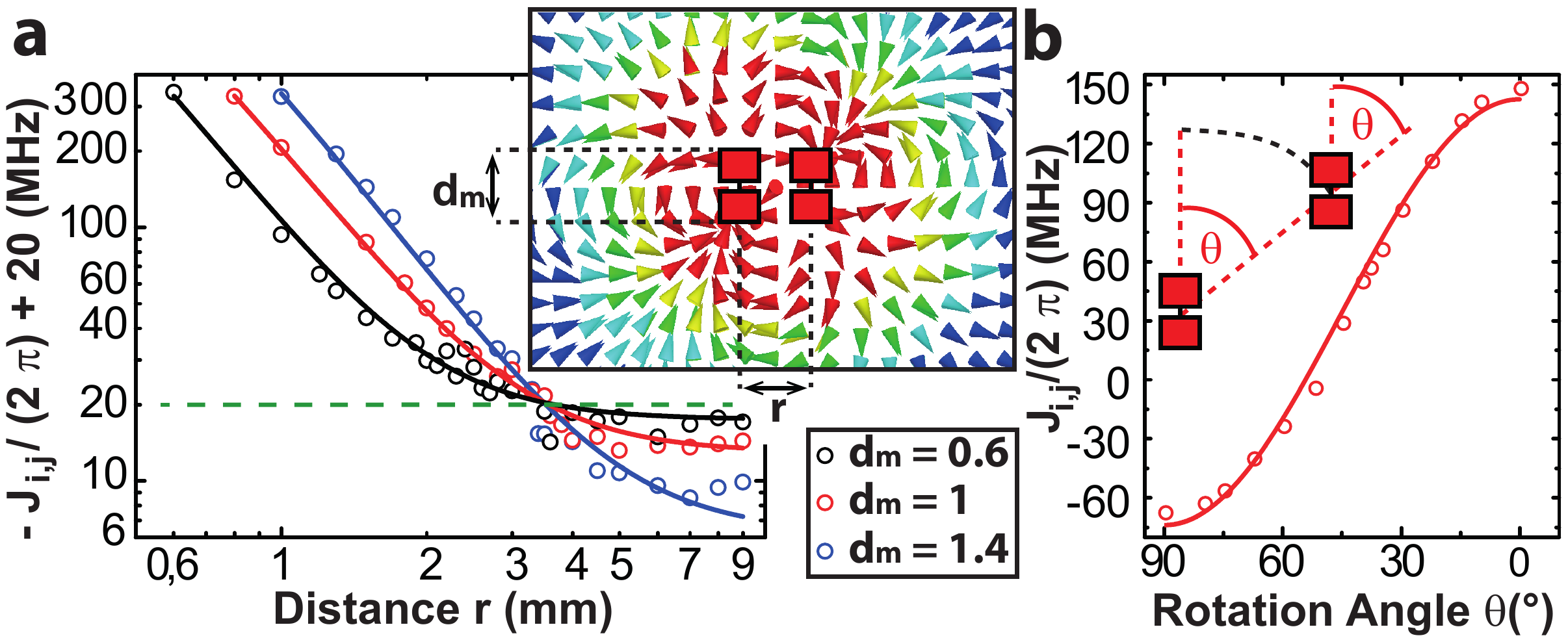}
\caption{Microscopic finite-element simulation (see text) of the dependence of the coupling $J_{i,j}$ between two qubits. Panel (a): distance dependence at $\theta_i=\theta_j = 90^\circ$ for three different antenna length. The dashed line indicates $J_{i,j}=0$. The inset displays a typical result from HFSS simulations~\cite{Ansys}, with the dipoles surrounded by the computed electric field (intensity decreasing from red to blue).
Panel (b): angular dependence illustrated for the case $\theta_i=\theta_j=\theta$ at distance $r=1.5$~mm. The solid line plots Eq.~\ref{dipole} with $J_0,r_m$ extracted from best fits in (a).}
\label{interaction} 
\end{figure}

The system dynamics we are interested in, is described by a generalized XY Hamiltonian of the form
\begin{equation}\label{eqH}
H /\hbar= \sum_{i,j}\frac{J'(\theta_i,\theta_{j})}{|{\bf r}_{ij}|^3}(S^+_iS^-_j + h.c.)+ \sum_j h_j S^z_j 
\end{equation} 
where $S_j^\alpha$ are spin operators at the lattice site $j$, ${\bf r}_{ij}$ is the distance vector between $i$ and $j$, the last term describes a disordered transverse field, and the inter-qubit couplings $J_{ij} = \frac{J'(\theta_i,\theta_{j})}{|{\bf r}_{ij}|^3}$ are in frequency units. The key element of our implementation is the realization of different patterns of quantum frustration~\cite{lacroix2011introduction} by tuning the form of the interaction couplings (Fig.~\ref{interaction}). As discussed below, the latter display a rich dependence as a function of the dipole angles $\theta_j$ (see Fig.~\ref{interaction}b): this dependence is essential to tune the coupling between different spins from positive to negative, or to (approximately) set it to 0. 
Even more crucially, the $1/|{\bf r}_{ij}|^3$ dependence (see Fig.~\ref{interaction}a) can be exploited to further modify the magnitudes of the couplings. 
In the theory of 1D and ladder systems~\cite{bosonization,Giamarchi2003}, where the dipolar interaction is effectively local, this allows independent tunability of nearest-neighbor (NN) and next-nearest-neighbor (NNN) exchange, a fundamental ingredient to realize bond-order solids~\cite{Majumdar1969,Mishra2013}. We focus specifically on this case below, showing how, within our proposal, such states of matter are robust against both disorder and qubit decoherence. The setup can be straightforwardly extended to 2D geometries, where qualitatively new features emerge due to the non-locality of the dipolar couplings: in particular, the dynamics of Eq.~\eqref{eqH} can by-pass the Mermin-Wagner-Hohenberg theorem~\cite{mermin1966,Hohenberg1967}, and thus support phases of matter with true long-range order even at finite temperature~\cite{peter2012}. We emphasize that all of these phenomena are directly accessible within our proposal thanks to the naturally occurring dipolar interactions, while such interactions would be challenging to be implemented via wiring.

\paragraph{Model system. -}
We now describe how the many-body dynamics of Eq.~\eqref{eqH} can be realized in our setup, using state-of-the-art circuit QED technology, with the design flexibility and long coherence times $1/\kappa$ ($\kappa\leq 2 \pi \times 100$~kHz) of Transmon qubits ~\cite{Paik:2011,Rigetti:2012}. This is schematically illustrated in Fig.~\ref{fig_system}. Several Transmon qubits (in red) fabricated on a piece of sapphire (light blue area) are mounted inside a waveguide cavity (grey box). The Transmon qubits can be fabricated in an essentially arbitrary lattice configuration with locally controllable orientation. The waveguide cavity around the qubits is used to readout the state of selected qubits and apply a drive to a subset of qubits, providing means for both adiabatic state preparation and probing (see next section). This selectivity can be again achieved by partially rotating the qubits inside the cavity by a few degrees, such that their dipole moment has a finite overlap with the electric field in the cavity. To add more flexibility and mitigate disorder, each qubit can be fabricated on an individual piece of sapphire, and can be probed at room temperature~\cite{Gloos2000}. For simplicity these individual pieces are not shown in Fig.~\ref{fig_system}. From fabrication one can expect a disorder of about 3-4\% in the Josephson energy which one should be able to reduced to less than 1\% (around $2 \pi \times 30$~MHz spreading in $h_j$) by preselecting the qubits according to their measured normal state resistance.

\paragraph{Designing the interaction. -}
The unique feature of our approach is the design flexibility in the interqubit interaction which results from the dipole-antenna structure of the Transmon. As one can expect, two dipole antennas in the near field interact like two magnetic spins~\cite{Jackson:1998}. By designing the shape and size of the antenna we can realize large interaction strengths ($J_{ij} \approx 2 \pi \times 100$~MHz) at inter-qubit distances of about 1 mm. The interaction between the qubits ultimately comes from an effective capacitance between the antennas, similar to \cite{Viehmann2013a, Viehmann2013b}, with an angle and distance dependence akin to magnetic dipoles. 

To confirm and quantitatively assess the inter-qubit dipolar interactions, we have performed a finite element study to determine the coupling strength and dependence on distance and angle. These simulations are carried out by {\it ab initio} numerically solving the 3D Maxwell equations on finite grids (HFSS software~\cite{Ansys}), and as such provide an extremely accurate quantitative benchmark for our modeled inter-qubit couplings. The quantitative insights gathered from the simulations are of basic importance to underpin the effects of the specific configuration that would be accessible in realistic setups, where the influence of the surrounding waveguide cavity and finite size effects of the Transmon antenna have to be systematically understood. A set of results can be seen in Fig.~\ref{interaction} (see also~\cite{supmat} for additional results). Assuming a dipole-dipole interaction with an additional component due to the dispersive coupling of both qubits to the cavity with strength $g$ (determined from independent simulations), we expect the following spatial dependence:
\begin{eqnarray}\label{dipole}
J_{1,2}= \frac{g^2 d_m^2}{2 \Delta}  \sin \phi_1 \sin \phi_2 - J_0 d_m^2 \frac{\cos \theta - 3 \cos \theta_1 \cos \theta_2}{(r-r_m)^3}.
\end{eqnarray}
The first term takes into account the qubit-qubit interaction mediated via the fundamental cavity mode with $\phi_1, \phi_2$ the orientation of the qubits relative to the orientation of the electric field in the cavity. Higher order cavity modes contribute with a similar term with a coupling $g'$ depending on the mode structure; as these are typically very far detuned, we neglect them in the following. The second term is the direct dipole-dipole interaction between the qubits with $r=|{\bf r}_{12}|$ being the distance between their centers. Both, the qubit-cavity interaction strength (which depends on the location of the qubit in the cavity) and the dipole-dipole interaction $J_0$ get modified by the length of the antenna, giving rise to the term $d_m$ in the Eq.~\ref{dipole}, with $d_m$ the normalized antenna length. The remaining variables in this equation are the qubit cavity detuning $\Delta\approx 2 \pi \times 1.5$~GHz and the angles $\theta_1, \theta_2 $. These are the angles formed by the two dipoles with respect to a line connecting their centers, as can be seen in Fig.~\ref{interaction}(b). $r_m$ corrects for finite size effects of the dipole antennas.

One can see in Fig.~\ref{interaction}(a) that the qubit-qubit interaction strength in our numerical simulations very closely follows the analytical expression Eq.~\ref{dipole}. The only fitting parameters in this expression are $J_0$, which is the same for all three datasets, and the $r_m$, which is adjusted for each length. The dependence on the distance for different antenna sizes can be seen in Fig.~\ref{interaction}(a). The interaction strength between the qubits for a parallel orientation is negative for short distances and falls of as $1/r^3$. For larger distances it becomes positive due to the additional coupling mediated off-resonantly via the cavity. This leads to the effect that qubits at a given distance (in the case of our cavity about 3.5~mm) do not interact with each other as these terms exactly cancel out. \footnote{Such a canceling of the direct inter-qubit interaction can be very useful for quantum information experiments where such an always-on interaction might not be desirable.} 

By rotating one qubit around the other, as shown in Fig.~\ref{interaction}(b), we can see that the interaction strength goes from a negative value for a parallel orientation to a twice as large positive value for collinear qubits, as expected for a dipole-dipole interaction. The analytical curves agree well with the numerical simulations and demonstrate that the spatial dependence of the interaction behaves like a magnetic dipole-dipole interaction. Furthermore we note that the cavity mediated term can be fully suppressed by rotating the qubits perpendicular to the electric field of the cavity mode, as shown schematically in Fig.~\ref{fig_system}. The finite-element simulations are in very good agreement with a simple analytical circuit model (see \cite{supmat}), which already captures the important interaction features.

\begin{figure}[t]
\begin{center}
\includegraphics[width = 0.48\columnwidth]{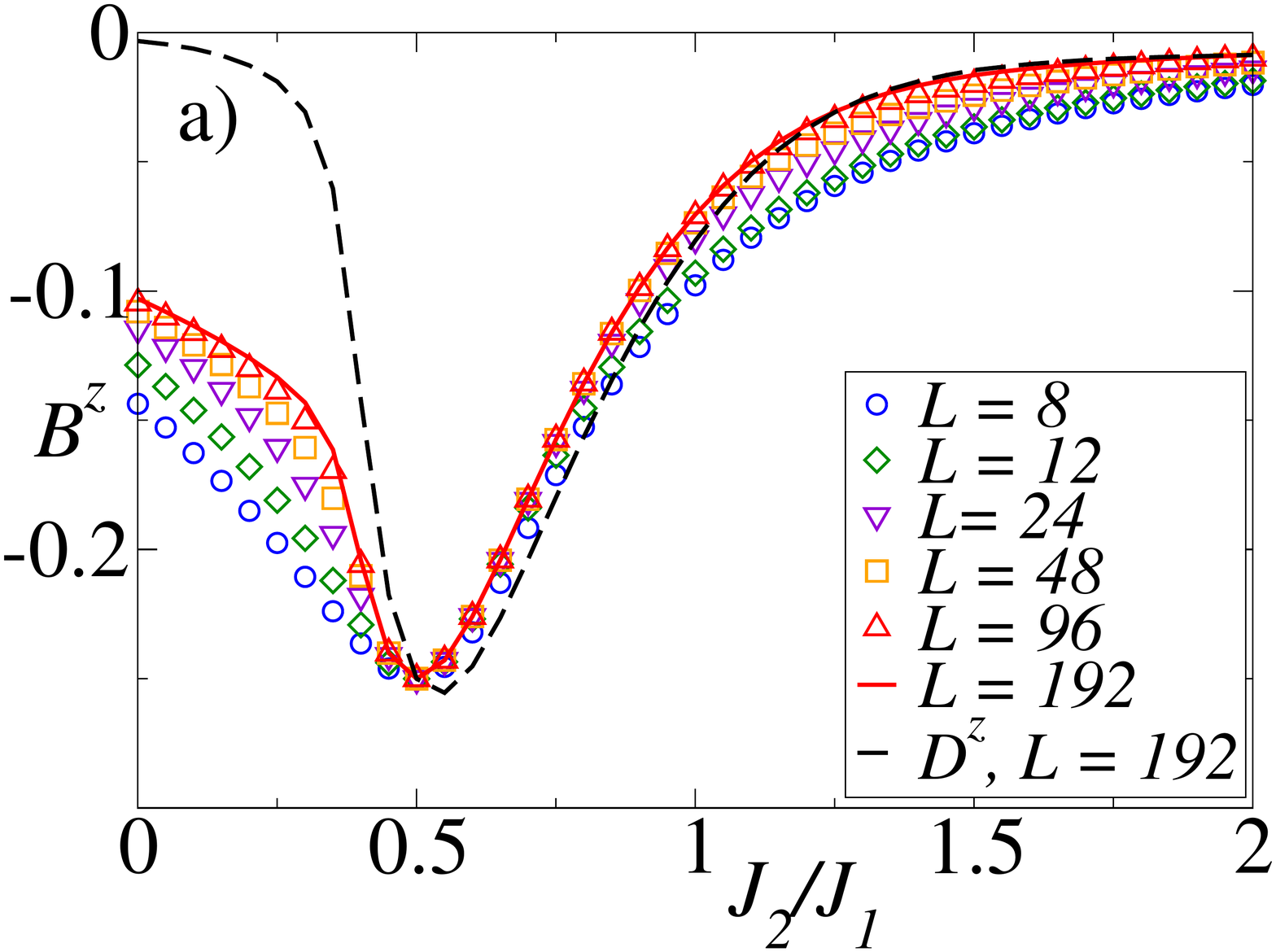}
\includegraphics[width = 0.48\columnwidth]{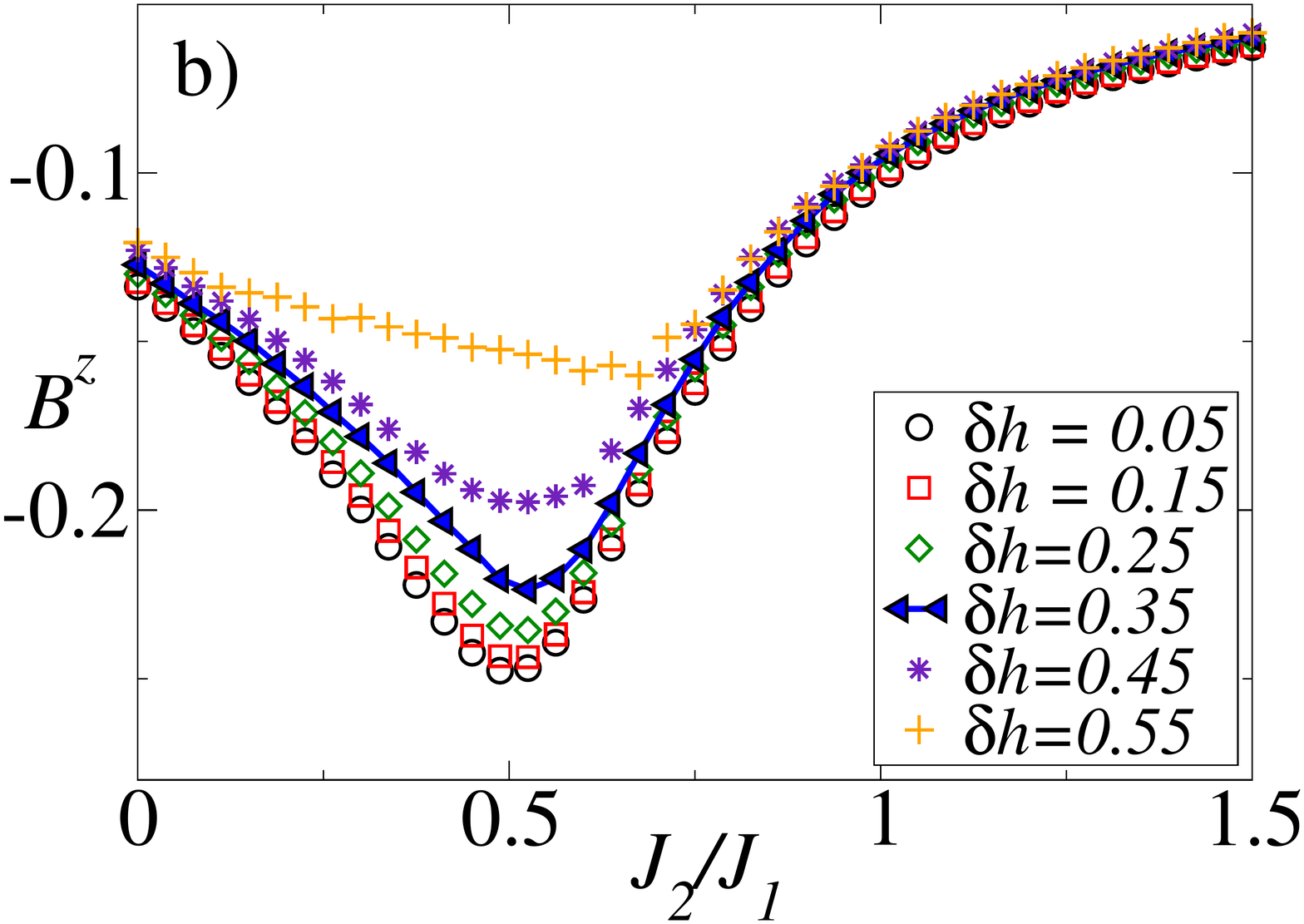}
\caption{Static properties of Eq.~\eqref{Ham}. Panel (a): Comparison between the BOP $D^z$ for a large system (L=192 spins, dashed black line), and the bond correlation in the middle of the chain for samples of different sizes. Panels (b): Effects of disorder on the dimer phase. Up to disorder Gaussian spreadings of order $\delta h\simeq 0.45J_1$, the bond correlation is weakly unaltered i.e. the phase is stable against local perturbations. 
The blue curve represents a conservative estimate for typical experimental setups. Here, $L=10$ sites, results show averaging of  500 disorder realizations per point. }
\label{fig_static} 
\end{center}
\end{figure}

\paragraph{Frustrated spin ladders and dimer phases. -}
The setup discussed above paves the way to the realization of a broad class of lattice spin models~\cite{lacroix2011introduction}. The key challenge on the theory side is to identify many-body phenomena which can be realized within this toolbox and, crucially, are stable against the most common imperfection sources such as disorder and decoherence. Here, we show how a archetypical state of matter, a valence bond solid akin to the Majumdar-Ghosh (MG) state in frustrated spin chains~\cite{lacroix2011introduction,Majumdar1969,Hikihara2001,Mishra2013}, can be realized and probed with currently available technology. Specifically, we are interested in the model in Eq.~\eqref{eqH} on a triangular ladder geometry (see Fig.~\ref{fig_system}b). To simplify our discussion, since dipolar interactions are effectively local in 1D, we consider only NN and NNN couplings, that is:
\begin{equation}\label{Ham}
H_{\textrm{lad}}/\hbar = \sum_{i}(J_1S^+_iS^-_{i+1} + J_2S^+_iS^-_{i+2} + \textrm{h.c.}). 
\end{equation}
This model~\cite{Mishra2013,Greschner2013} is known to exhibit a phase transition from a superfluid phase ($J_1\gg J_2$) to a dimer phase (DP) around $J_2/J_1\simeq 0.33$, which is stable up to relatively large couplings~\cite{Hikihara2001}. This DP can be easily understood at the MG point $J_2/J_1 = 0.5$, where the ground state factorizes onto a product of local triplets~\cite{Mishra2013}:
 \begin{equation}
|\Psi \rangle_{GS} = \prod_{j=1}^{L/2}\otimes (\mid\uparrow_{2j-1}\downarrow_{2j}\rangle+\mid\downarrow_{2j-1}\uparrow_{2j}\rangle)
\end{equation}
as illustrated in Fig.~\ref{fig_system}(b). Beyond being a textbook example of a valence bond solid in 1D, the DP has triggered further interest in connection to the physics of the Peierls instability in Hubbard models~\cite{Kumar2010}. 
In general, the DP is indicated by a bond-order parameter (BOP):
\begin{equation}
D^{\alpha} =\langle \sum_{j=1}^{L-1} D_j^\alpha\rangle, \quad D_j^\alpha  = (-1)^j S^\alpha_jS^\alpha_{j+1},\;\alpha=x,z
\end{equation}
which indicates the spontaneous formation of triplets along the odd-even bonds on an open chain. Deep into the DP, the phase can be identified in an even simpler way by monitoring the bond correlation of a single pair of spins in the middle of the chain, $B^z = D^z_{L/2}$ (for $L = 4n, n\in\mathbb{N}$)\footnote{The same conclusions hold for $D^x_{L/2}$.}. In Fig.~\ref{fig_static}(a), we compare the behavior of the dimer correlation $B^z$ (red line) against the BOP (dashed line) as a function of the coupling ratio $J_2/J_1$ for a chain of $L=192$ sites, obtained using density-matrix-renormalization group simulations~\cite{DMRG1}~\footnote{All simulations are carried out using open boundary conditions to reflect the experimental setup. For the DMRG simulations, we used up to $L=512$ states per block
and $8$ finite-size sweeps in chains up to $L = 288$ sites, resulting in a truncation error  $<10^{-7}$ on the ground state energy.}. Both quantities are maximized in the DP, as expected; in addition, close to the MG point, the difference between the BOP and $B^z$ is only at the few percent level even for very small system sizes of $L=8$ spins, providing an ideal observable to detect the DP even in relatively small samples.  Measurement techniques to access these observables exploiting current technology are discussed in Ref.~\cite{supmat}.

A crucial point is, to which extent the DP and the corresponding signatures on small size systems are stable against disorder. Qualitatively, we expect that, close to the MG point, the phase is stable against weak disorder for $h_j<J_2$, as breaking a single triplet will cost energy $\simeq J_2$. We have performed an exact diagonalization study of the resilience against disorder, which is included as a Gaussian distribution of $h_j$ around 0 with spread $\delta h $ ($J_1$ setting the energy unit). In Fig.~\ref{fig_static}b, we show the 
behavior of the bond correlation as a function of $J_2/J_1$ for a system of $L=10$ sites and 
different disorder strength and average values. Up to a spread of around $0.45$, the signatures of the dimer phase are very stable. Given that typical disorder strengths would be $\delta h\simeq 0.2/0.3$, this makes the observation of the DP realistic with current fabrication technology. 

\begin{figure}[t]
\begin{center}
\includegraphics[width = 0.98\columnwidth]{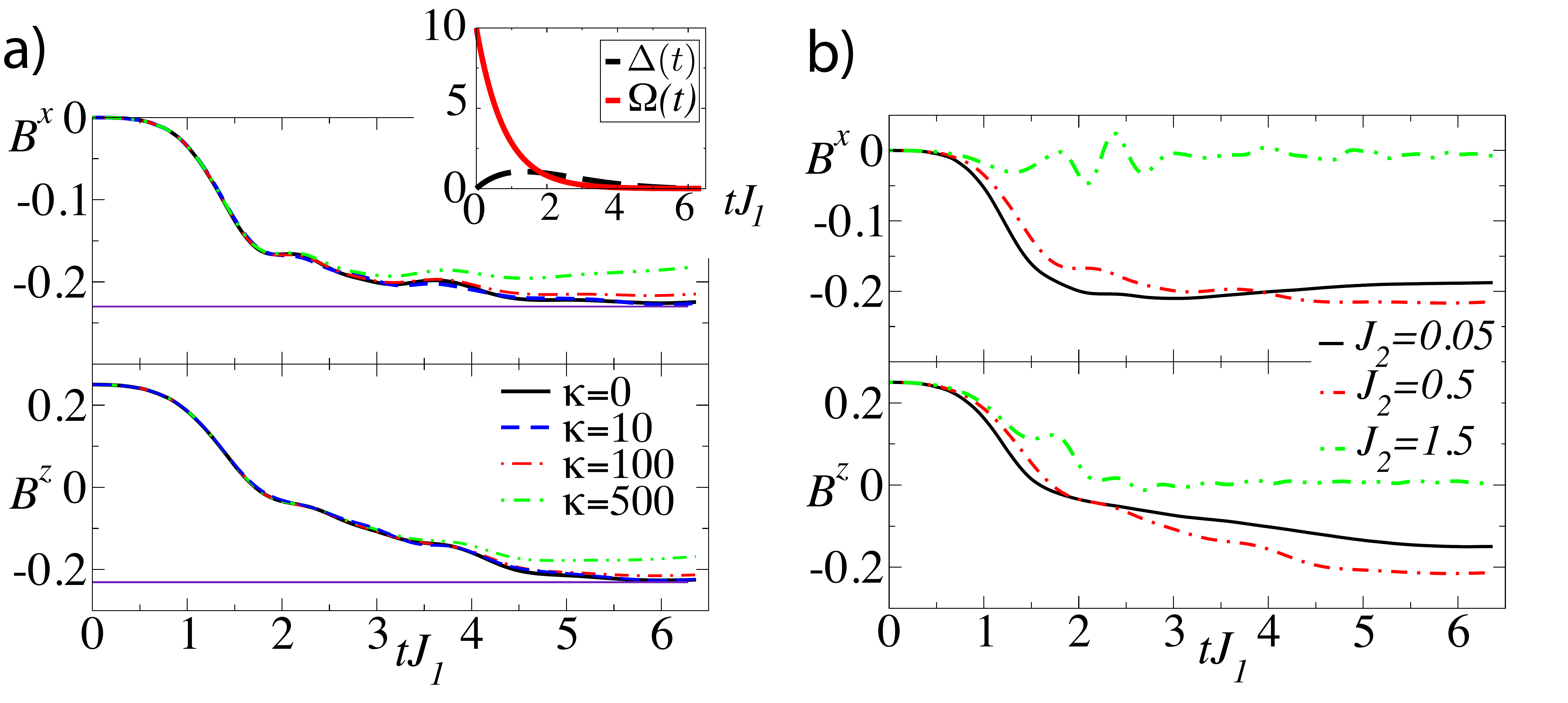}
\caption{Adiabatic state preparation. Panel (a): state preparation of the dimer phase for a system with $L=6$, $2J_2=J_1=2\pi\times100$ MHz, and different decay rates (in $2\pi\times$ kHz units), with $\gamma=\kappa$. The upper (lower) panel denotes the bond $x$ ($z$) correlation averaged over 2000 disorder realizations, with $\delta h /J_1= 0.25$. The straight lines denote the ideal case. Up to decoherence rates of order 100 kHz, the final observables at the end of the ramp are within 2\% of the expect one. The inset shows how the ramping parameters (in units of $J_1$) evolve with time. Panel (b): same as in (a) for fixed $\kappa=2\pi\times100$ kHz, and different values of $J_2$ (in units of $J_1$). Outside of the DP (red dot-dashed line), the order parameters have very different final values, as the final state is not dimerized.}
\label{fig_sramp1} 
\end{center}
\end{figure}

\paragraph{Adiabatic state preparation and effect of dissipation. -} We finally discuss how the DP
can be realized in realistic settings via adiabatic state preparation. The system is initially in the vacuum state, where no excitations are present, that is, $|\Psi\rangle_{in} = \prod_j\otimes\mid\downarrow\rangle_j$\footnote{Possible single excitations in the system due to thermal effects do not significantly affect the dynamics}. An external drive is used to inject excitations, and can be modeled by a time-dependent Hamiltonian $H_{d} = \sum_j\Omega_j S^x_j + \sum_{j}\Delta_j(t)S^z_j$. The dynamics is well described by a master equation in Lindblad form~\cite{Gardiner2000}:
\begin{eqnarray}\label{masterEq}
\dot{\rho}(t) & = & -\frac{i}{\hbar} [H_T(t), \rho(t)]+ \\
&+& \frac{1}{2}\sum_{\ell = 1, 2L} [2C_j\rho(t)C^\dagger_j - \rho(t)C^\dagger_jC_j - C^\dagger_jC_j\rho(t)]\nonumber
\end{eqnarray}
where $H_T(t) = H + H_d(t)$, the jump operators $C_{j} = \sqrt{\kappa}_jS^+_j$ and $C_{j+L} = \sqrt{\gamma}_jS^z_j$ describe single-spin decay and dephasing, respectively, with constants $\kappa, \gamma\lesssim 100$~kHz. In our ramps, we employed a staggered driving field, $\Omega_j = ((1+(-1)^j) + \delta_{j, L/2})\Omega$, which is provided by the local spin addressability using angle tuning (see Fig.~\ref{fig_system}(a)), and performs significantly better than other choices. The specific shape of the sweep parameters we choose is illustrated in the inset of Fig.~\ref{fig_sramp1}a: first, the detunings $\Delta_j$ are slowly switched off while ramping up $\Omega$. Subsequently, the driving is switched off in order to reach the desired final state~\footnote{We note that, during the ramp, additional dipolar terms of the form $(S^+_iS^+_j+\textrm{h.c.})$ can also become resonant. We have numerically verified that such terms do not qualitatively affect the adiabatic loading.}.

We quantitatively address the full system dynamics including disorder and decoherence by numerically solving Eq.~\ref{masterEq} for systems up to $L=10$ spins. In Fig.~\ref{fig_sramp1} (a), we plot the average magnetization per site (blue dashed) and the bond correlation (both in the $x$- and $z$-basis, black and red line, respectively) as a function of time, with a system including decoherence rates with $\kappa=\gamma = 2\pi \times 10, 100, 500$~kHz, and a disorder strength $\delta h =2\pi \times 25$~MHz. The effect of decoherence along the ramp is minimal: the final order parameters deviate only a few percent from the expected value even for $\kappa=2\pi\times100$ kHz, making the phase detectable for typical experimental conditions. For values of the ratio $J_2/J_1$ outside of the DP, dimerization does not occur (Fig.~\ref{fig_sramp1} (b)). Quantum optimal control techniques~\cite{Doria:2011rz,Motzoi:2011oq} could eventually be employed to further improve the ground state preparation.

\paragraph{Conclusions and outlook.  -} 
We have shown how to use the naturally occurring dipolar interactions in 3D superconducting circuits to realize a platform for analogue quantum simulation of XY spin models. The possibility of realizing arbitrary lattice geometries with locally-tunable dipole moments, in combination with their large interaction strength, opens the door to the investigation of a series of phenomena in quantum magnetism in both 1D~\cite{Hyman1996,batista2014} and 2D~\cite{peter2012}, complementing the remarkable developments in cold atom and trapped ion systems~\cite{Bloch:2012, Blatt:2012,Georgescu:2014kq}. Our ideas are not limited to Transmon qubits, but could be implemented with, e.g., Xmon qubits~\cite{Barends:2014} or Fluxonium qubits coupled to an antenna~\cite{Vool:2014}. It would be interesting to explore these developments in view of realizing Hamiltonian dynamics for surface code architectures~\cite{Fowler:2012} or as a building block for coupled cavity array experiments~\cite{Hartmann:2006,Gerace:2009}.

\paragraph{Acknowledgments. - }

M.D. thanks Fabio Ortolani for help with the DMRG code. Part of the time-dependent simulations have been based on QuTip libraries~\cite{qutip}. We want to thank Simon Nigg and Michael Hatridge for feedback on the manuscript. This work is partially supported by the ERC Synergy Grant UQUAM, SIQS, and SFB FoQuS (FWF Project No.~F4006-N16)

 %%%%%
 %%%%%
 %%%%%
 %%%%%

% supplementary material here 

 %%%%%
 %%%%%
 %%%%%
 %%%%%

\newpage
\onecolumngrid

\begin{center}{
\bf \Large Supplemental Material to \\'Dipolar Spin Models with Arrays of Superconducting Qubits'
}
\end{center}

\vspace*{0.5cm}
{\center{
\hspace*{0.1\columnwidth}\begin{minipage}[c]{0.8\columnwidth}
In the first part and second part of this appendix the details of numerical finite element simulations and the lumped element circuit model are explained. In the third part the Hamiltonian for two capacitively coupled Transmon qubits inside a cavity is derived. More details on the qubit state measurement and flux tuning are given in part four. In part five, we present an extended discussion of the dimer order parameters utilized in the main text to characterize the dimer phase.\end{minipage}
}
}

\vspace*{0.5cm}

\twocolumngrid

\section{\label{sec:HFSS}HFSS simulations}
\begin{figure}[b]
	\centering
	\includegraphics[width=0.9\linewidth]{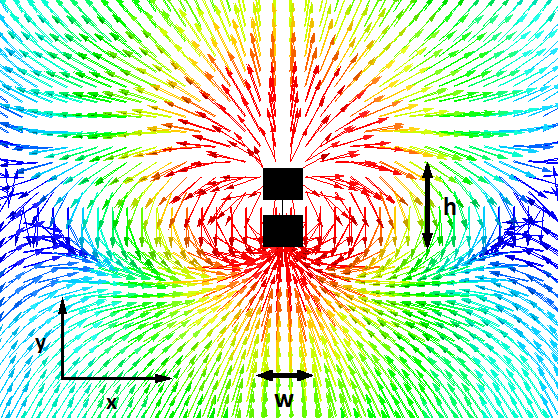}
	\caption[Short]{Electric field vectors of the qubit mode plotted on the sapphire substrate with the qubit at the center. A rainbow color code is used to show the field intensity (with red being high intensity).}
	\label{fig:Vectorfieldonsinglequbit}
\end{figure}

In this section, finite element electromagnetic simulations that were performed using the ANSYS HFSS~\cite{SAnsys} software are explained in more detail. Qubits are modeled as linear LC resonators (see also~\cite{SNigg2012}). The capacitance of the resonator originates from the capacitance between two rectangular films that are separated by distance $d$. These pads couple to electromagnetic waves like a dipole antenna. The Josephson junction of the qubit is modeled by a lumped element inductance connecting the two antenna pads. Fig.~\ref{fig:Vectorfieldonsinglequbit} shows the electric field distribution on a sapphire substrate that holds the qubit structure together with structure variable names used in this note.   

\begin{figure}[b]
	\centering
	\includegraphics[width=0.9\linewidth]{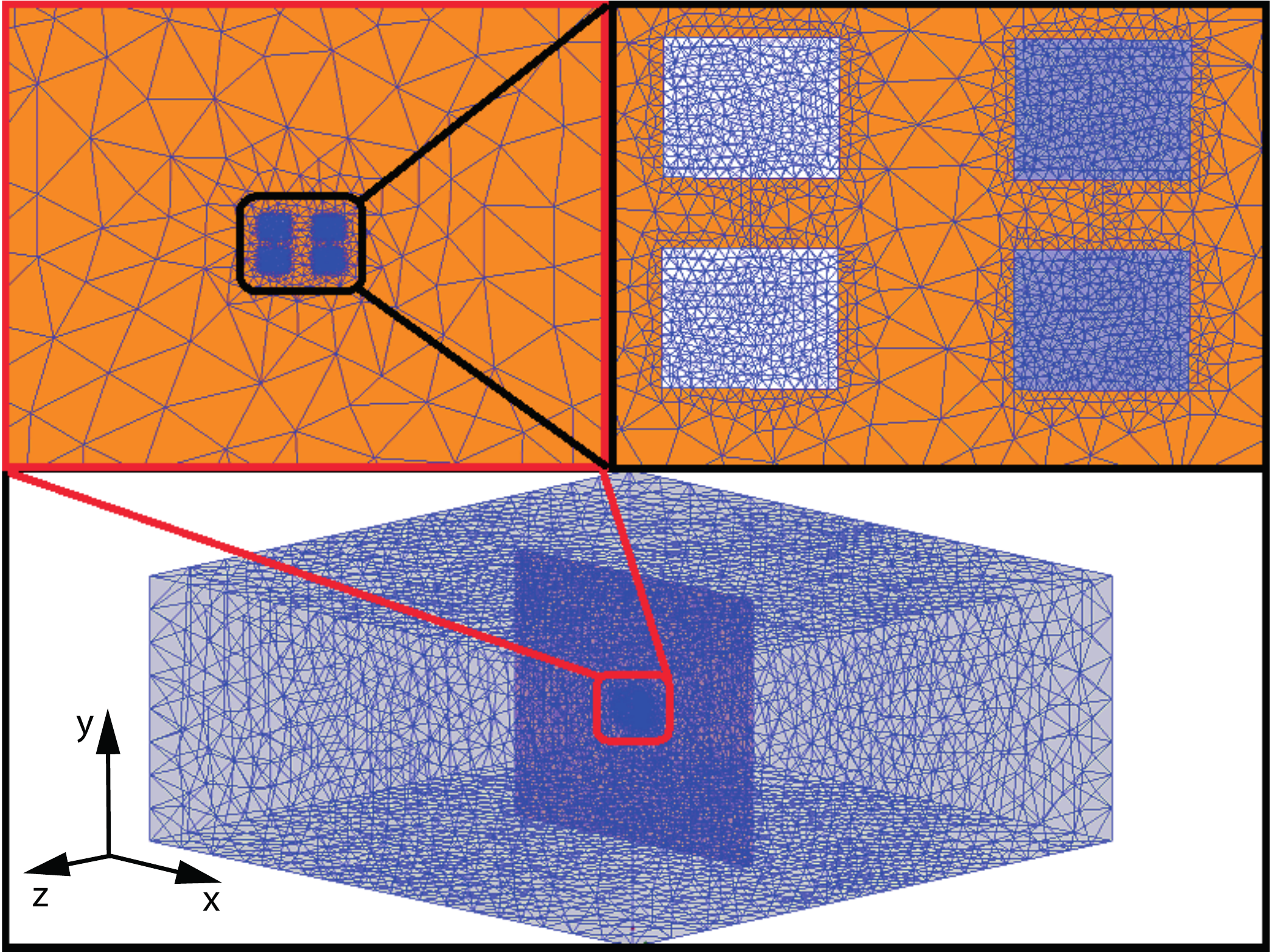}
	\caption[Short]{Mesh used to calculate the electromagnetic fields around the qubits. The left half of the image shows the meshing needed for the cavity that contains a sapphire substrate with two qubits on it. The right half shows meshing around one of the qubits. The coordinate system used in this paper is defined in the lower left.}
	\label{fig:Meshes}
\end{figure}

It should be noted that finite-element calculations of a structure with objects that are orders of magnitude different in dimensions requires plenty of computational resources. In this work, extremely fine meshing is needed in areas where accurate understanding of the physics of small objects (like qubits) is needed. Fig.~\ref{fig:Meshes} shows how different mesh sizes are used for different structure size to accurately calculate the eigenmodes of the electromagnetic fields of two qubits in a cavity.

In this study, the coupling strength is defined as the minimum difference in the mode frequencies of two qubits when the inductance of one qubit is swept in order to tune its mode frequency in and out of resonances with the other(see Fig. \ref{fig:modes}).

\subsection{\label{singleQubit}Single qubit inside cavity}
\begin{figure}[]
\centering
\includegraphics[width=1\linewidth]{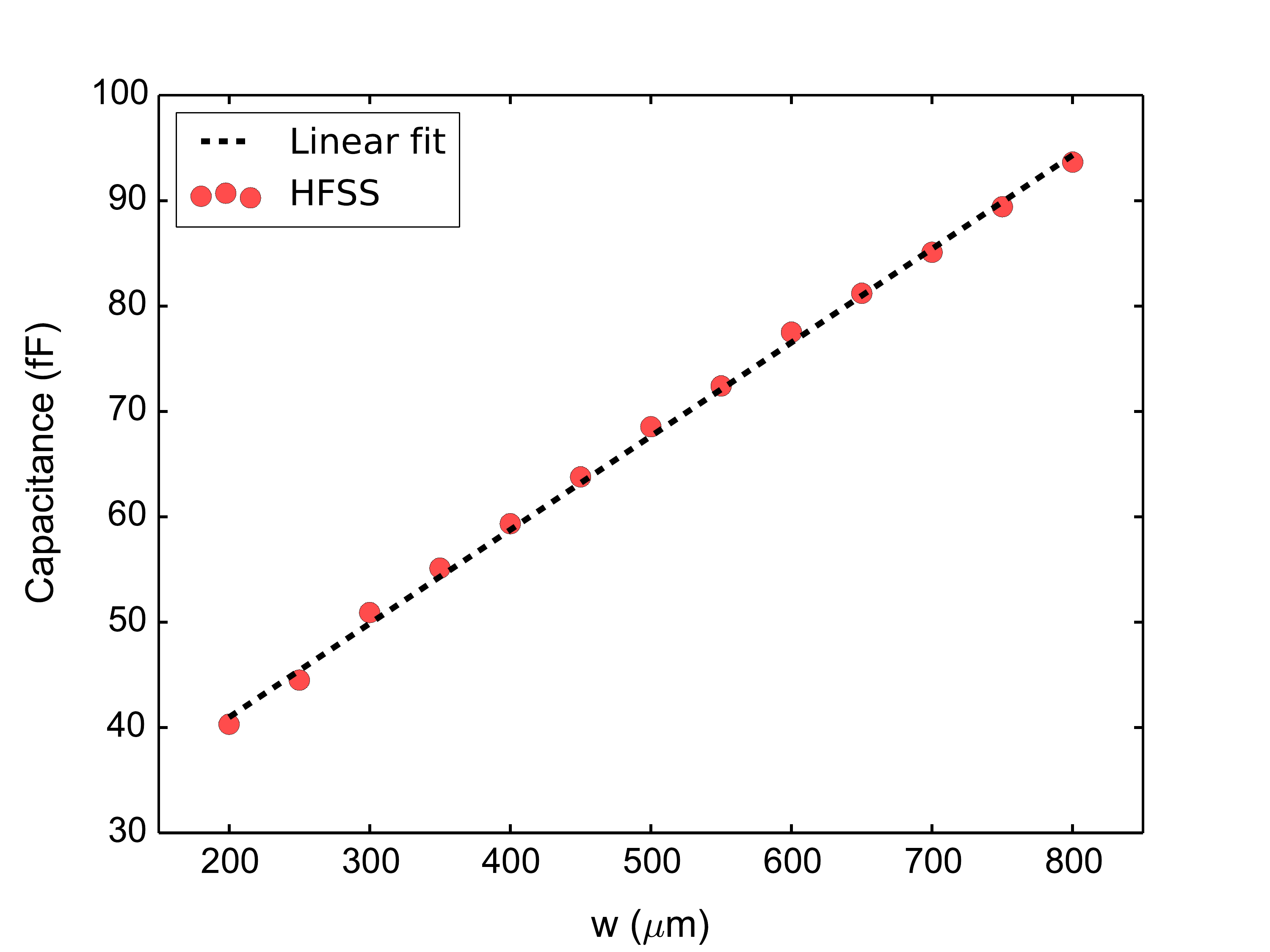}
\caption[C vs. width]{Extracted capacitance of the qubit as a function of the width of the antenna paddles. The dashed line is a linear fit to the data.}
\label{fig:capacitance}
\end{figure}

In order to better understand the effect of the antenna geometry on the qubit properties, a few cases have been studied. First, the width of the antenna pads $w$ was changed and the corresponding capacitance was extracted as shown in Fig.~\ref{fig:capacitance}. It can be concluded that the capacitance increases linearly as a function of the pad width.

\begin{figure}[]
\centering
\includegraphics[width=1\linewidth]{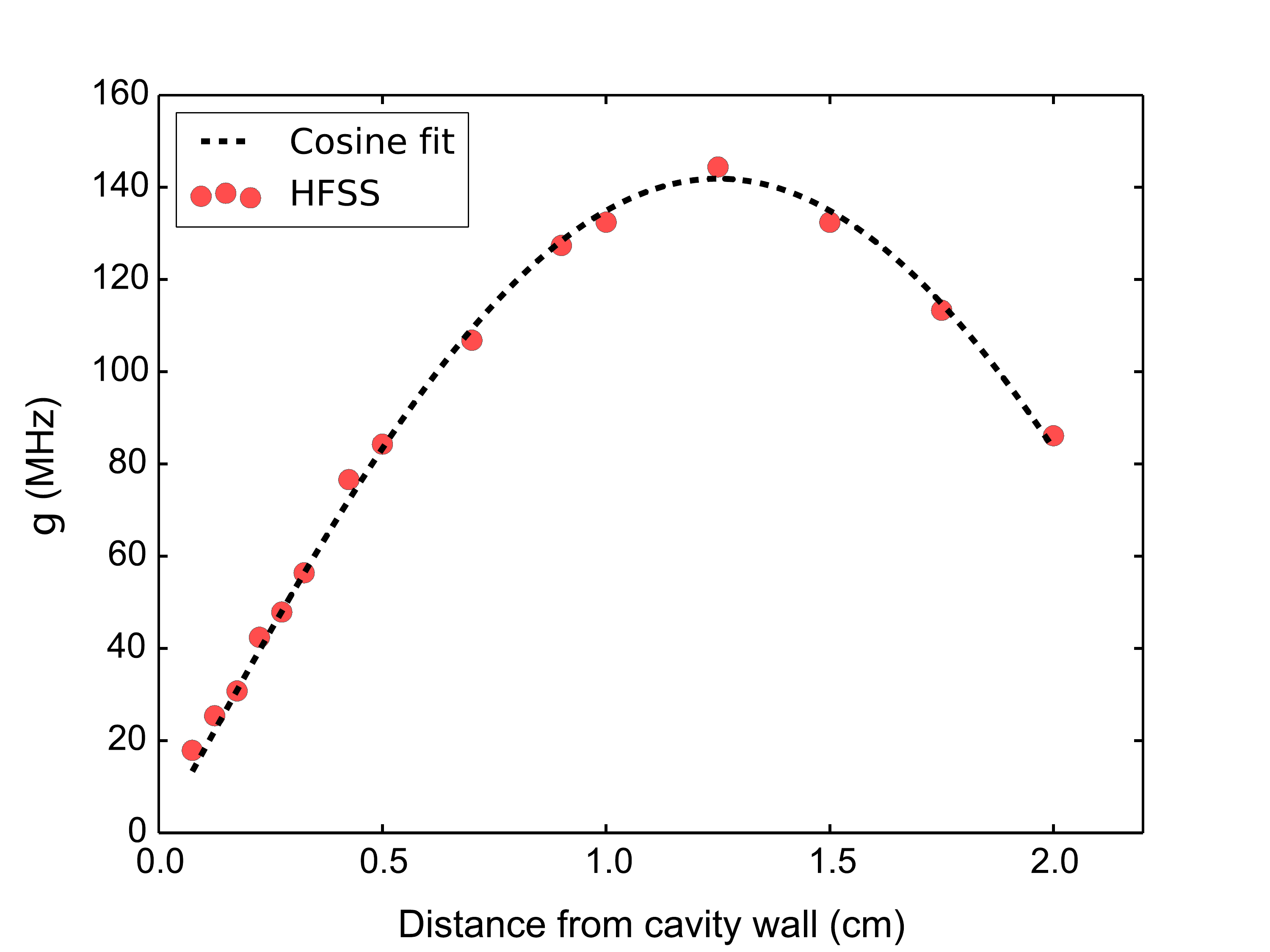}
\caption[$J_{i,j}$ vs. position in cavity]{Coupling strength $g$ of the qubit as a function of its position inside the cavity. The qubit was placed in the center of the cavity in the y and z direction and was moved along the x axis. The dashed line is a cosine fit to the data.}
\label{fig:cosine}
\end{figure}

Then, the coupling strength of the qubit to the cavity $g$ was calculated as a function of position of the qubit with respect to the side wall of the cavity. The coupling strength has a cosine form as expected from a standing wave in the cavity. The results are shown in Fig.~\ref{fig:cosine}.

\begin{figure}[]
\includegraphics[width=1\linewidth]{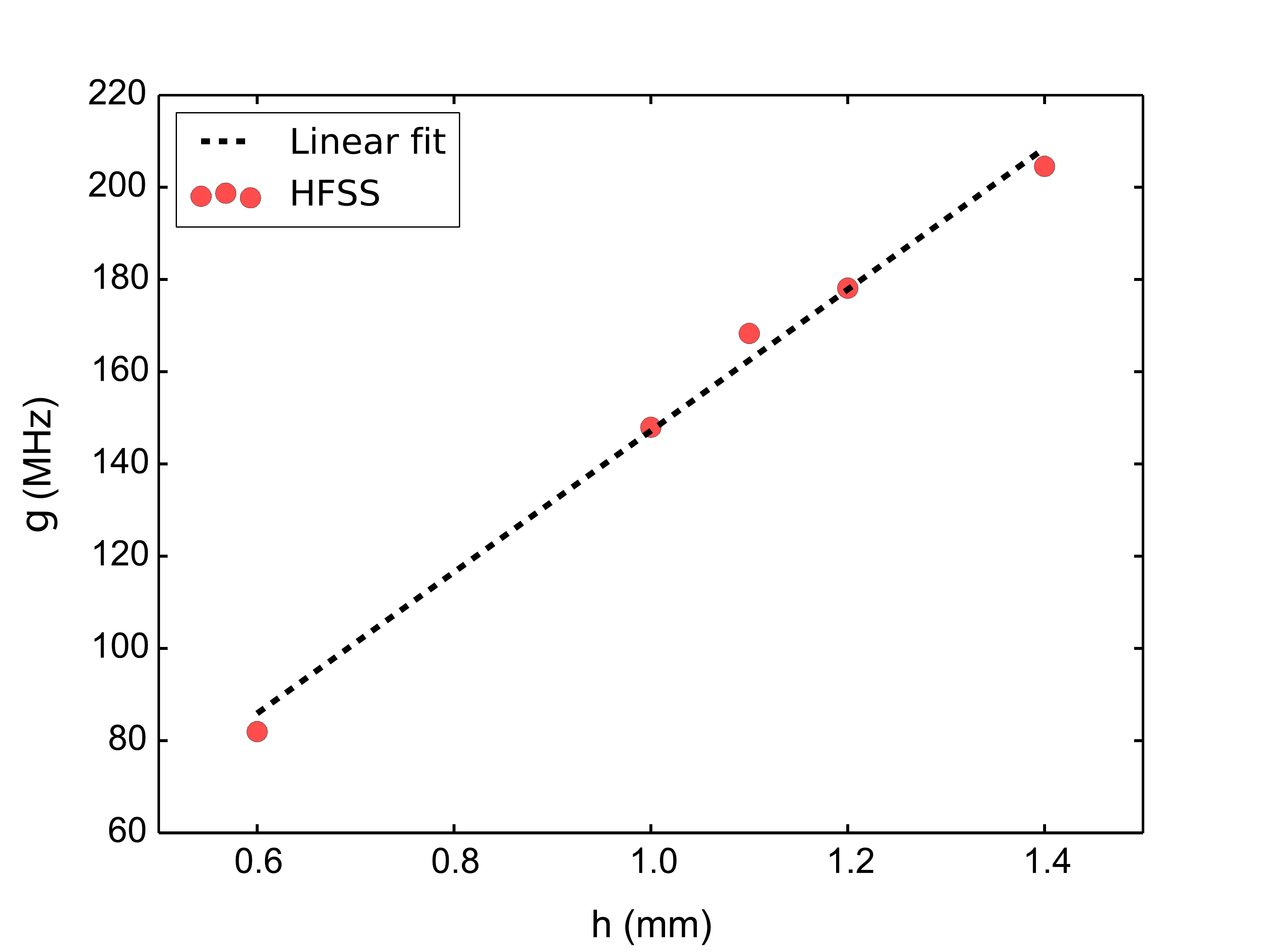}
\caption[g vs. height]{Coupling $g$ of the qubit to the cavity as a function of its antenna length. Since a change in antenna length will also change the resonance frequency, the width of the antenna has been adjusted to keep the resonance frequency constant. The dashed line is a linear fit to the data.}
\label{fig:height}
\end{figure}

The coupling strength $g$ was also studied as a function of the pad height $h$ as shown in Fig.~\ref{fig:height} where the qubit was located at the center of the cavity. It can be seen from the figure that the coupling strength increases linearly as a function of the antenna height. This is in agreement with dipole antenna physics.  
 
\subsection{\label{twoQubits}Two qubits in a cavity}
\begin{figure}
\centering
\includegraphics[width=1\linewidth]{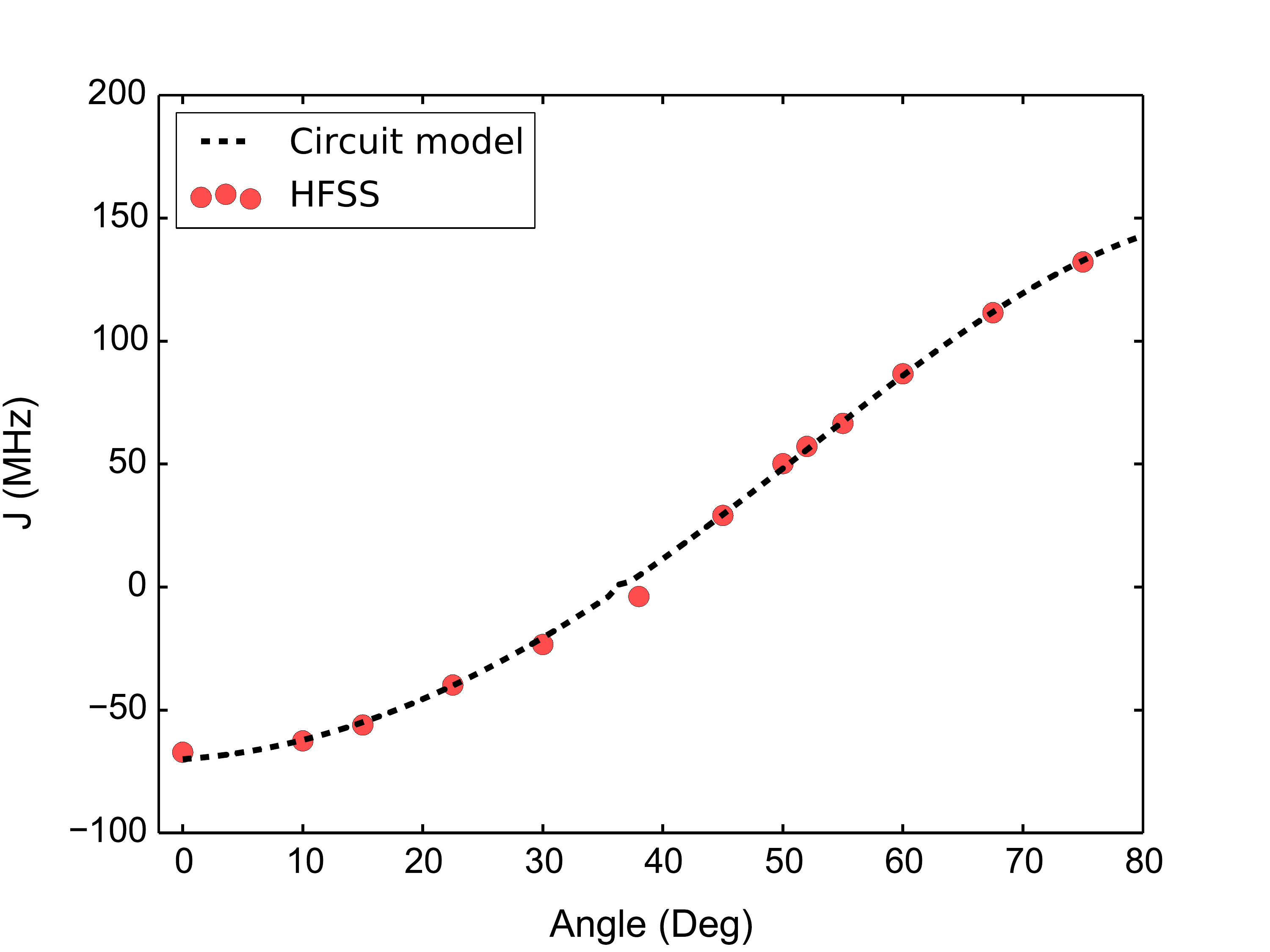}
\caption[$J_{i,j}$ vs. angle]{Coupling $J_{i,j}$ as a function of center-center vector angle. The two antennas are parallel and their center to center distance was kept at 1.5~mm. The dashed line shows a fit to the circuit model (see section~\ref{sec:CircMod}).}
\label{fig:theta_12}
\end{figure}

Fig.~\ref{fig:theta_12} shows how the coupling strength varies when one of the qubits is kept at a constant distance (center-to-center) of 1.5~mm from the other while the angle of the vector that connects the centres of two qubits is changed. In our simulations, the coupling strength is defined to be a positive number. However, in this particular case, the sign of the coupling strength has been changed to negative for values below $\approx$ 40 degrees. This is to highlight the fact that the sign of the coupling is different in the two regions. This sign change can be seen from HFSS simulations which show parallel/anti-parallel field vectors for the lowest mode, indicating a positive/negative coupling. As explained in section~\ref{sec:CircMod}, the results are in strong agreement with the proposed circuit model. 
 
\begin{figure}
\centering
\includegraphics[width=1\linewidth]{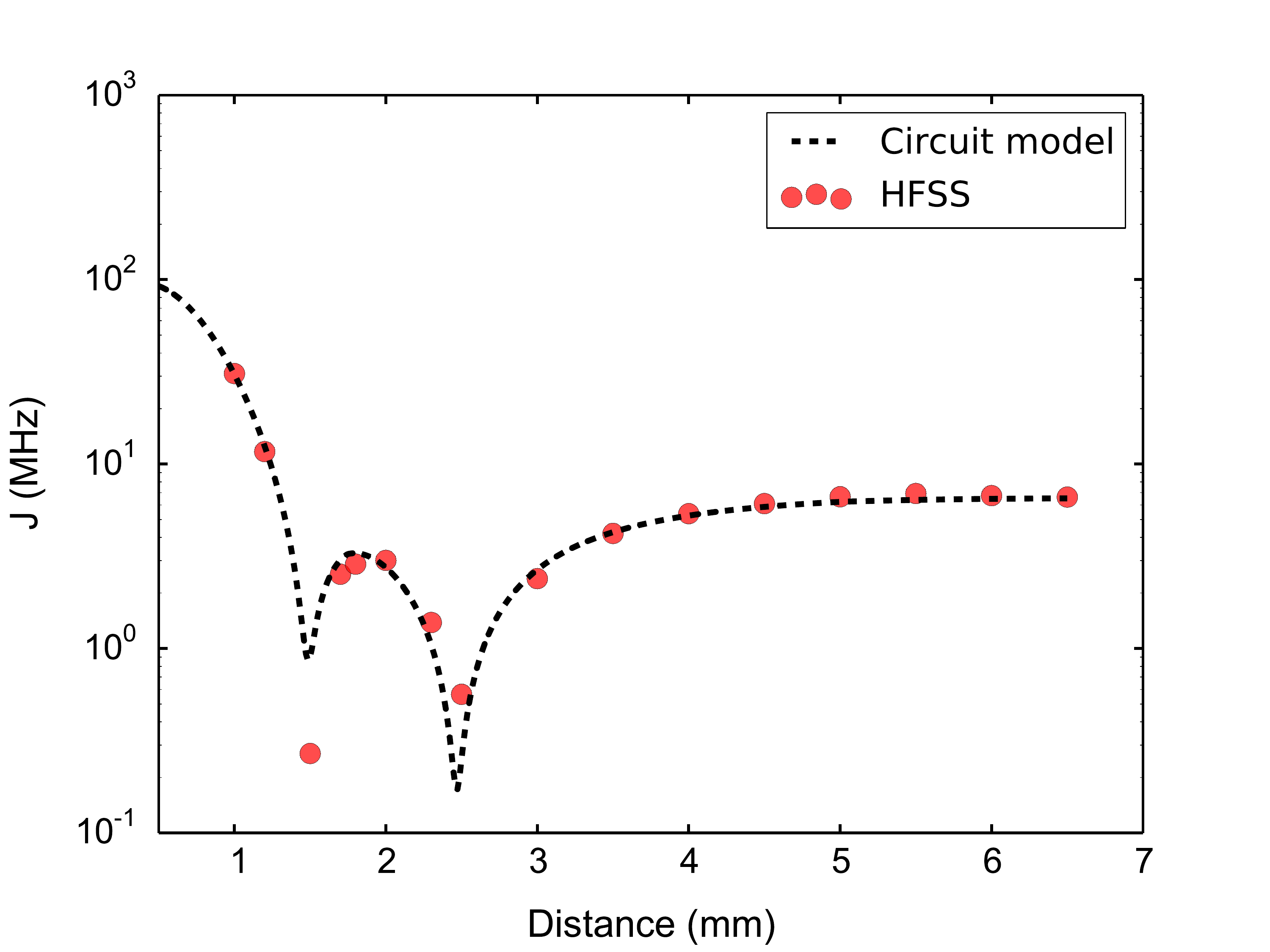}
\caption[$J_{i,j}$ vs. distance]{Coupling $J$ as a function of distance change along x axis. The center to center separation along the Y axis is kept at 1~mm and one of the qubits is moved along the x axis with respect to the other. The results suggest double suppression of coupling in agreement with the circuit model (dashed curve). }
\label{fig:Cutline}
\end{figure}

In another study, the coupling strength was evaluated when moving one qubit with respect to the other along a straight line. In Fig.~\ref{fig:Cutline}, the coupling strength is shown for a qubit moving along the x axis while the distance along y is kept at 1~mm. This geometry is inspired by the circuit model prediction that there should be two zeros along this line. A fit to the circuit model is also shown in the graph. 

\section{\label{sec:CircMod}Circuit model}
In this section, details of a circuit model that represents two-qubits inside a cavity are explained. The model is used to evaluate the coupling strength between the elements. 

\subsection{\label{subsec:CirDiag}Circuit diagram}
\begin{figure}[]
	\centering
	\includegraphics[width=1\linewidth]{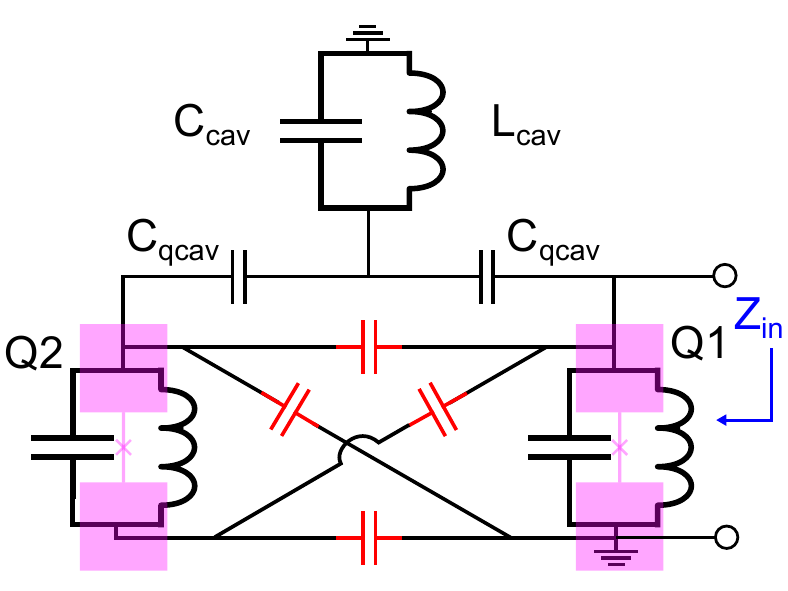}
	\caption[Circuit diagram]{Circuit model corresponding to two qubits in a cavity. Values of the four capacitors drawn in red decay as $1/r^2$ where $r$ is the center to center distance between pads of the antenna. This models the dipole-dipole coupling of two qubits. The two purple structures on Q1 and Q2 represent the physical shapes of the qubits. }
	\label{fig:circuit}
\end{figure}

Dissipation-less LC resonators are used to model the cavity and qubits (see also~\cite{SNigg2012}). The nonlinearity of the qubits is not taken into account as it does not affect the coupling strength. Capacitors are used in order to account for the coupling between elements. The coupling between each qubit and the cavity is represented by a coupling capacitor $C_{cav}$. As qubits are assumed to couple to each other through their dipole antenna, a total of four capacitors are used to represent coupling of each antenna pad to the two pads of the other qubit (see Fig. \ref{fig:circuit}).   

\subsection{\label{subsec:Coupling}Coupling strength}
In order to calculate the coupling between elements, the following method is used:
The impedance of the system ($Z_{in}$) seen through one of the qubits (here \q{1}) is calculated. The poles of the impedance are the resonance modes of the total system, in this example three. By sweeping the inductance $L_1$ (of \q{1}) we can see the avoided crossing between the modes (see Fig. \ref{fig:modes}). Coupling is defined as the minimum distance between the two modes at the avoided crossing.  

\begin{figure}[]
	\centering
	\includegraphics[width=1\linewidth]{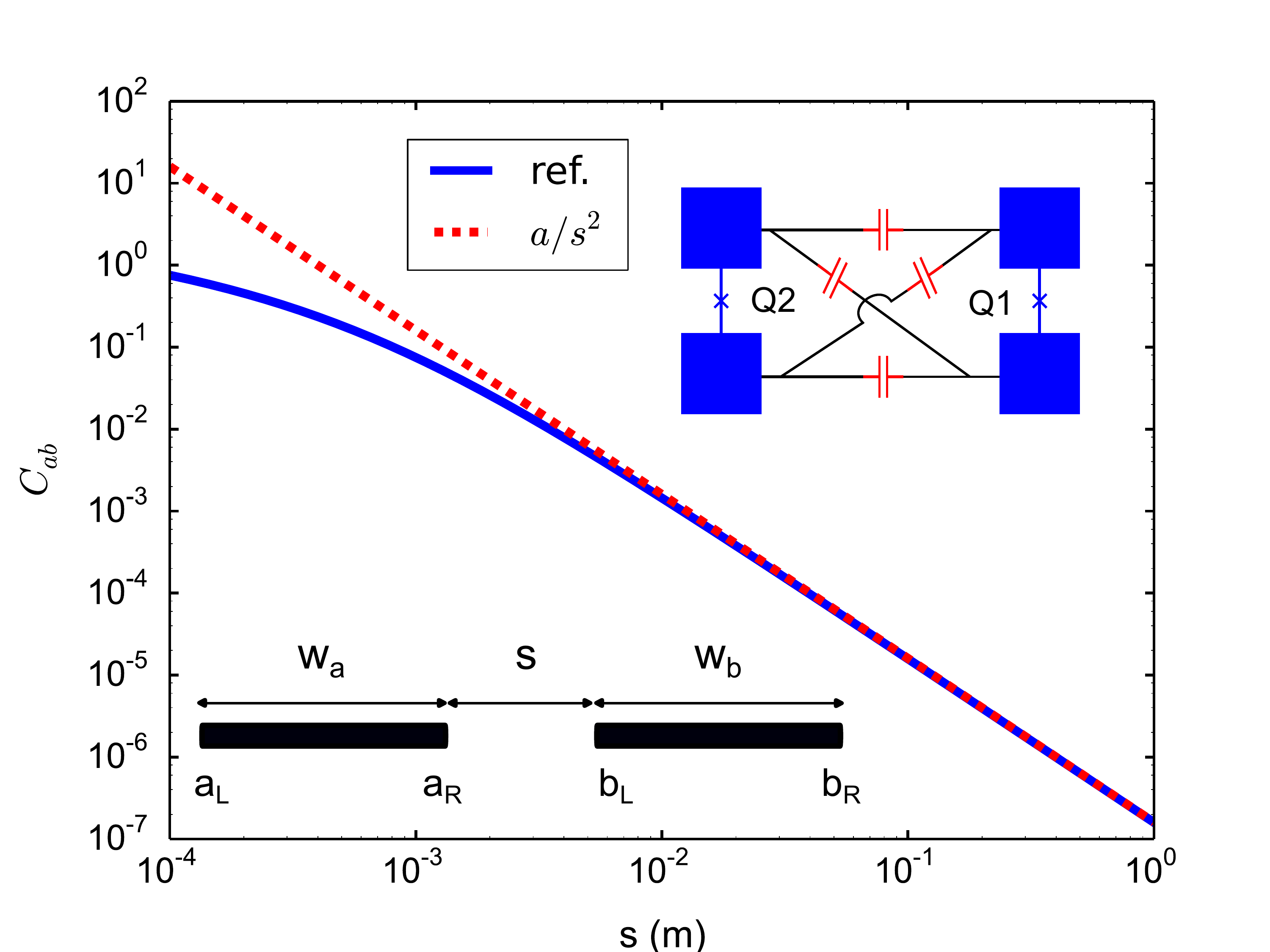}
	\caption[Circuit diagram]{Capacitance between two co-planar paddles (solid line) as a function of the separation $s$ according to equation \ref{Eq:CapDecay}. The corresponding structure is shown at the bottom. The dashed line shows the quadratic decay used in this study to estimate the coupling capacitance. The structure at top right shows which capacitors in the circuit model follow the quadratic decay. }
	\label{fig:capacitance_decay}
\end{figure}

\begin{figure}
\centering
\includegraphics[width=1\linewidth]{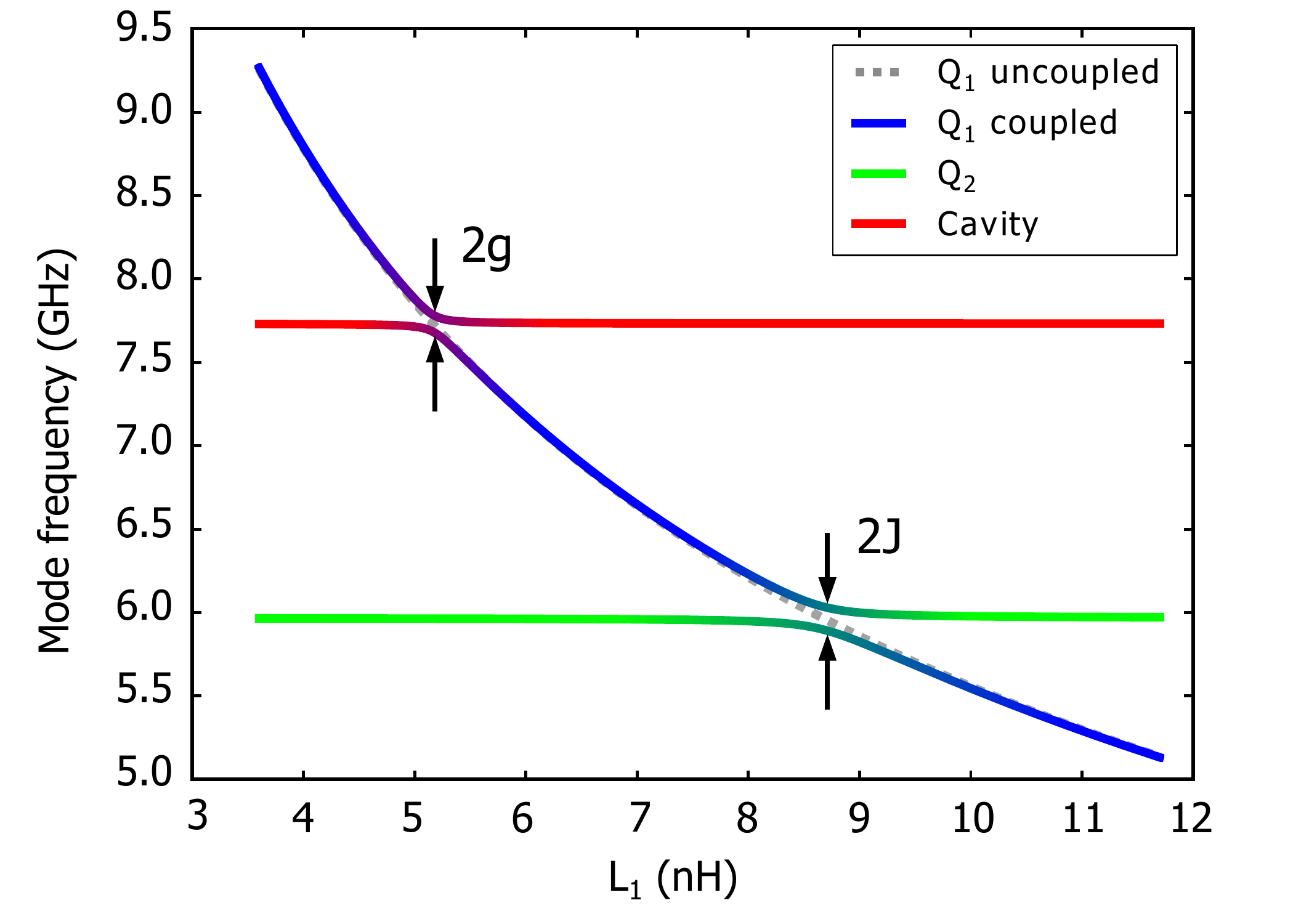}
\caption[Avoided crossings.]{Example for three modes of a system with two qubits and a cavity.The minimum distance of two modes is defined as the coupling $2J_{i,j}$ if it is two qubits and as $2g$ if it is the qubit cavity coupling. Note that despite the fact that \q{2} has 9~nH of inductance, the crossing happens at lower inductances due to the coupling to \q{1}. Dashed line shows how resonance frequency of a typical LC circuit would behave without coupling to the other resonators.}
\label{fig:modes}
\end{figure}

A further step was taken by relating the inter-qubit coupling capacitances (as in Fig.~\ref{fig:circuit}) to the physical distances among the qubit antenna poles. This distance dependence is assumed to be of the following form~\cite{SMartinis2014}:
\begin{equation}\label{Eq:CapDecay}
\begin{aligned}
C_{ab}&=(2\epsilon/\pi)\ln{\frac{(a_L-b_L)(a_R-b_R)}{(a_R-b_L)(a_L-b_R)}}\\
&=(2\epsilon/\pi)\ln{\frac{(w+s)^2}{(w+s)^2-w^2}}
\end{aligned}
\end{equation}
where $\epsilon$ is the average over dielectric constants of the substrate and the air (or vacuum),  $w=w_a=w_b$ is the width of antenna paddles and $s$ is the separation between closest edges as shown in the inset of Fig.~\ref{fig:capacitance_decay}. The parameters $a_L, a_R, b_R, b_L$ are defined in Fig.~\ref{fig:capacitance_decay}.

As this study concerns square shaped paddles (see top right inset of Fig.~\ref{fig:capacitance_decay}), it is not possible to apply equation \ref{Eq:CapDecay} to any arbitrary relative position of the two qubits without complicated corrections. Therefore, just to demonstrate the general behavior, a simplified relation (Eq.~\ref{Eq:CapDecayQuad} follows Eq.~\ref{Eq:CapDecay} when the separation $s$ becomes much larger than the paddle width $w$ - this effect is shown in Fig.~\ref{fig:capacitance_decay}) for the coupling is chosen which assumes that the capacitance decays quadratically as a function of the distance: 
\begin{equation}\label{Eq:CapDecayQuad}
C_{ab}=\frac{a}{s^2}
\end{equation}
where $a$ is a coefficient used to adjust the coupling strength. 

This means that the coupling strength $J_{i,j}$ can be calculated based on the physical position of the two qubits with respect to each other. In this study, \q{1} is assumed to be fixed in position and \q{2} changes its distance and angle with respect to \q{1}. 

%\subsubsection{\label{subsubsec:CplDist}Coupling as a function of distance}

\begin{figure}[]
	\centering
	\includegraphics[width=1\linewidth]{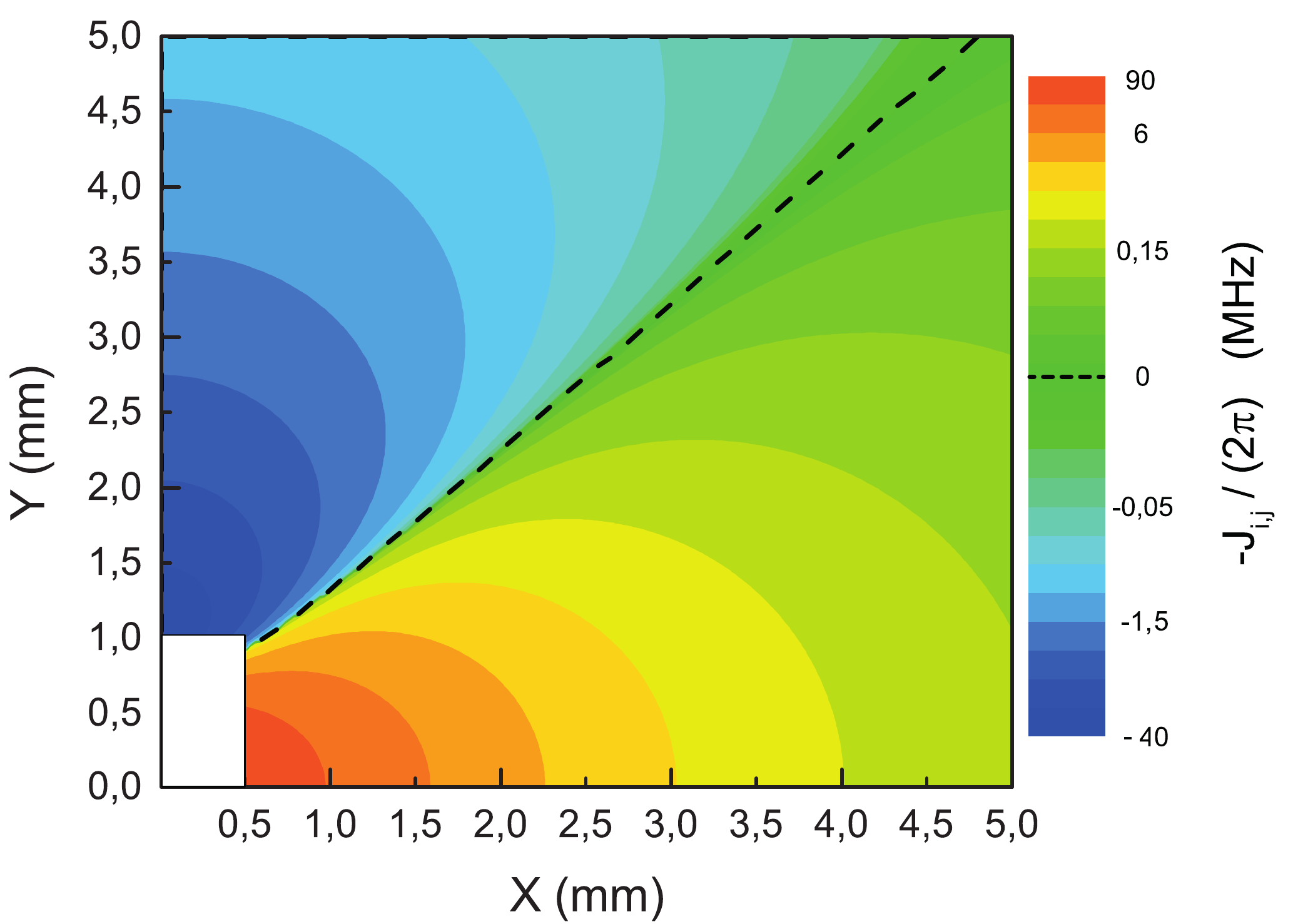}
	\caption[Coupling strength without a cavity.]{Contour plot showing the coupling strength between two qubits $J_{i,j}$ as a function of the distance between them, in the x and y direction, in absence of the cavity. The color scale is logarithmic to highlight the rapid change in $J_{i,j}$. In such a case, the dipole physics dominates and the coupling will be suppressed along a straight line. The white area is not accessible due to the dimensions of the qubits.}
	\label{fig:g_2D_nocav}
\end{figure}

\begin{figure}[]
	\centering
	\includegraphics[width=1\linewidth]{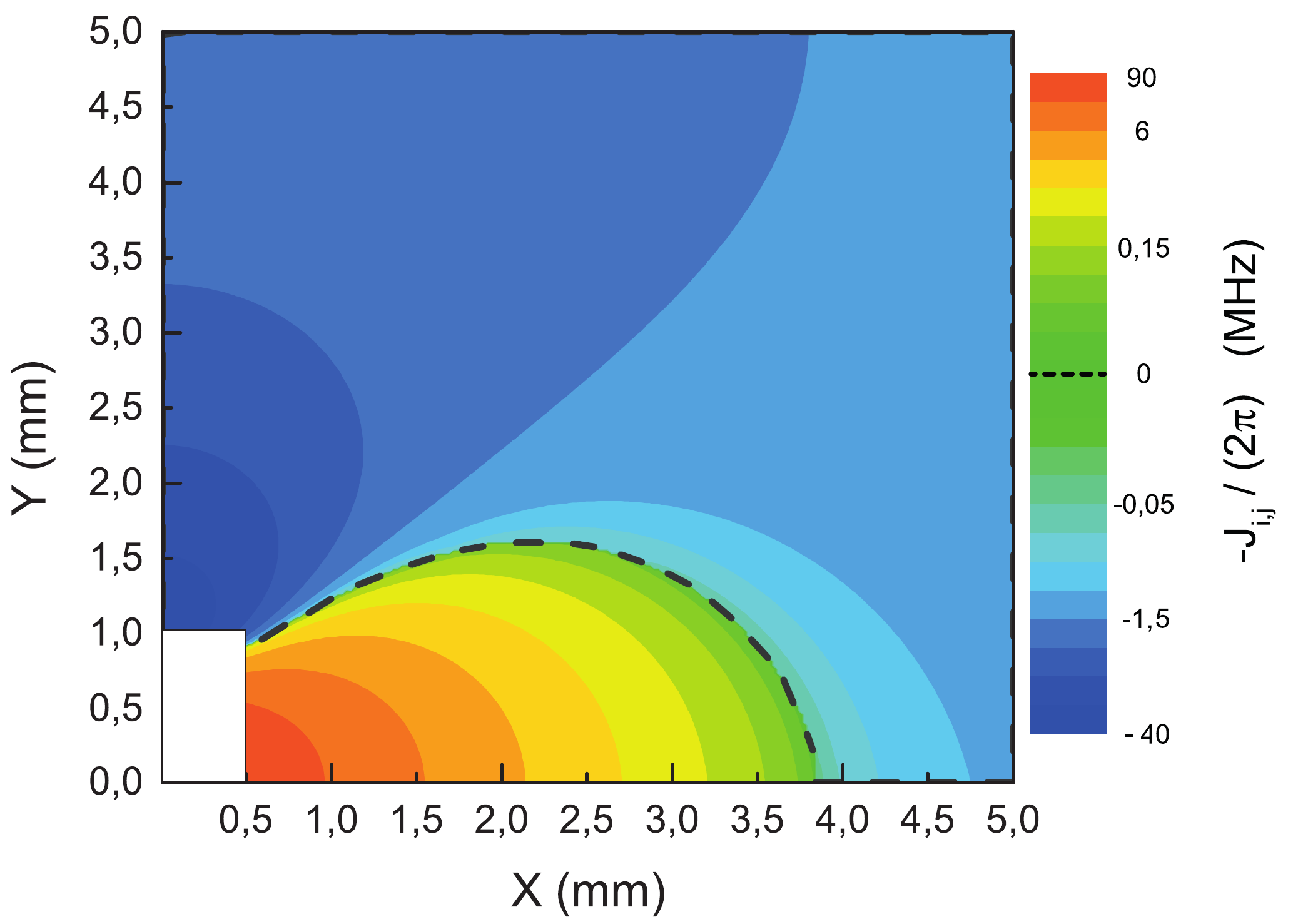}
	\caption[Coupling strength with a cavity.]{Contour plot showing the coupling strength between two qubits $J_{i,j}$ as a function of the distance between them, in the x and y direction, in presence of the cavity. The color scale is logarithmic to highlight the rapid change in $J_{i,j}$. The cavity distorts the straight line of suppressed coupling opening new possibilities for engineering of the coupling between two qubits. The white area is not accessible due to the dimensions of the qubits.}
	\label{fig:g_2D}
\end{figure}

Assuming that the two qubits are placed parallel with respect to each other, we have calculated the coupling $J_{i,j}$ as a function of the center-to-center distance. In Fig.~\ref{fig:g_2D_nocav} the results are shown for two qubits without a mediating cavity. As can be seen, the coupling is heavily suppressed along a straight line. This line can be thought of as when the dipole electric field from \q{1} is horizontal, that is perpendicular to the dipole antenna of \q{2} and hence, there will be no coupling. This is typical behavior of dipoles and is well expected. However, by adding a cavity to the model (see Fig. \ref{fig:g_2D}), one can observe that the straight line shape of suppression will no more hold and instead, a curved line emerges.

%\begin{figure}[]
	%\centering
	%\includegraphics[width=0.9\linewidth]{Figures/TwoEigenModes.png}
	%\caption[Different regimes of coupling]{Electric current directions for the lowest eigenmode for positive and negative '$J$' as explained in section \ref{twoQubits}. Each regime is marked by a different color range in figures \ref{fig:g_2D_nocav} and \ref{fig:g_2D}.}
	%\label{fig:antiparallel}
%\end{figure}

Adding the cavity has certain consequences on how two qubits couple. For example, if one moves \q{2} along the x-axis while it is held at $y=1$~mm (or $y=-1$~mm), the coupling might be suppressed twice. This is verified by HFSS simulations as seen in Fig.~\ref{fig:Cutline}. Our lumped-element model shows behavior that is consistent with HFSS simulations as already demonstrated in Figs.~\ref{fig:theta_12} and~\ref{fig:Cutline}.

\section{Hamiltonian derivation}
\subsection{Interacting Transmons}

\begin{figure}[t]
  \begin{center}
    \includegraphics[width=0.95\columnwidth]{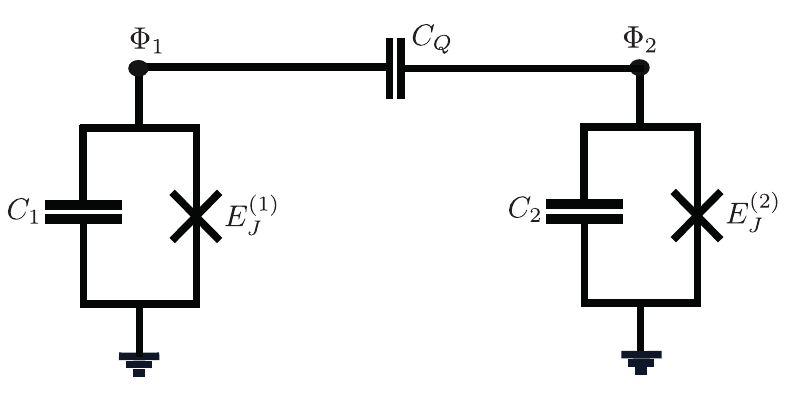}
  \end{center}  
  \caption{
Lumped element circuit model for two capacitively coupled superconducting qubits.
}
  \label{Fig1}
\end{figure}

Let us consider two interacting transmons as depicted in Fig.~\ref{Fig1}. The Lagrangian describing this system is given by

\begin{eqnarray}
L &=& \sum_{i=1,2} \frac{1}{2} C_{i} \dot\Phi_i^2 + \frac{1}{2} C_Q (\dot\Phi_1-\dot\Phi_2)^2 + \\ \nonumber
&+& \sum_{i=1,2} E_J^{(i)} \cos \left( \frac{\Phi_i}{\Phi_0} \right)
\end{eqnarray}
with $\Phi_0 \equiv \hbar/(2e)$. The Hamiltonian, $H = \sum_{i=1,2} Q_i \dot\Phi_i - L$ can then be written as
\beq
H = \frac{1}{2} \vec{Q}_T {\cal C}^{-1} \vec{Q} - \sum_{i=1,2} E_J^{(i)} \cos \left( \frac{\Phi_i}{\Phi_0} \right),
\eeq
with $\vec{\Phi}_T \equiv (\Phi_1,\Phi_2)$. For small phase fluctuations, we can expand the cosine and get $H=H_0+H_1$, with
\beq
H_0 &=& \frac{1}{2} \vec{Q}_T {\cal C}^{-1} \vec{Q} + \frac{1}{2} \vec{\Phi}_T {\cal L}^{-1} \vec{\Phi} \label{H0} \\ 
H_1 &\approx& - \sum_{i=1,2} \frac{E_J^{(i)}}{24\Phi_0^4} \Phi_i^4, \label{H1}
\eeq
where
\beq
 \mathcal{C}&=&\nonumber
\left(\begin{array}{cc}
C_1 +C_Q  & - C_Q \\
-C_Q & C_2 +C_Q \\
\end{array}\right), \\
 \mathcal{C}^{-1}&=& \frac{1}{D}\nonumber
\left(\begin{array}{cc}
C_2 +C_Q  & C_Q \\
C_Q & C_1 +C_Q \\
\end{array}\right), \\
 \mathcal{L}^{-1} &=&\nonumber
\left(\begin{array}{cc}
1/\alpha_1 & 0 \\
0 & 1/\alpha_2 \\
\end{array}\right),
\eeq
$D\equiv C_1C_2 + C_1C_Q + C_2C_Q$ and $1/\alpha_i \equiv E_J^{(i)}/\Phi_0^2$. We now quantize this Hamiltonian using the usual ladder operator replacement for $Q_j$ and $\Phi_j$ which is 
\begin{eqnarray}
Q_j = i Q_j^{\rm ZPF} (a_j^{\dagger} - a_j), \Phi_j = \Phi_j^{\rm ZPF} (a_j^{\dagger} + a_j),
\end{eqnarray}
where $Q_j^{\rm ZPF} \equiv \sqrt{\hbar\omega_j C_j^{\rm eff}/2}$, $\Phi_j^{\rm ZPF} \equiv \sqrt{\hbar\omega_j \alpha_j/2}$, $C_1^{\rm eff} \equiv \frac{D}{C_2+C_Q}$, $C_2^{\rm eff} \equiv \frac{D}{C_1+C_Q}$. Here, $\omega_1$, $\omega_2$ are the two solutions of $|{\cal L}^{-1} - \omega^2 {\cal C}| = 0$. This quantization gives
\beq\label{Hamilt}
H \approx \sum_{i=1,2} \omega_i n_i -  \sum_{i=1,2} \Lambda_i n_i^2 + \lambda_{12} (a_1^{\dagger}a_2 + {\rm h.c.}),
\eeq
where $\Lambda_j \equiv \frac{E_J^{(j)}}{4} \left( \frac{\Phi_j^{\rm ZPF}}{\Phi_0} \right)^4$, $\lambda_{12} \equiv \frac{C_Q}{D} Q_1^{\rm ZPF}Q_2^{ZPF}$, and $n_j\equiv a_j^{\dagger}a_j$. Notice that this Hamiltonian has been written in a basis of local modes. Diagonalization of the quadratic part via a Bogoliubov transformation gives rise to a Hamiltonian of the form~\cite{SNigg2012}
\beq \label{diagH}
H \approx \sum_{i=1,2} \xi_i n_i - \sum_{i=1,2} \alpha_i n_i^2 - \chi_{12} n_1 n_2.
\eeq

\subsection{Interacting transmons inside a cavity}

\begin{figure}[t]
  \begin{center}
    \includegraphics[width=0.8\columnwidth]{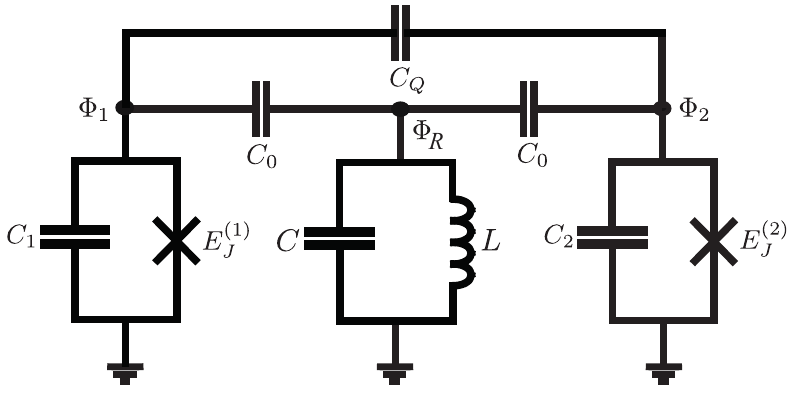}
  \end{center}  
  \caption{
Lumped element circuit for two interacting superconducting qubits in a cavity resonator.
}
  \label{Fig2}
\end{figure}

Let us now consider the circuit depicted in Fig.~\ref{Fig2} which is a simplified version of Fig.~\ref{fig:circuit} with the capacitive coupling network between the qubits reduced to one capacitor. This lumped element model will provide the right Hamiltonian but does not capture the angle and distance dependence explicitly. We first write the Lagrangian of the system, which reads
\beq
L &=& \sum_{i=1,2} \frac{1}{2} C_{i} \dot\Phi_i^2 + \sum_{i=1,2} E_J^{(i)} \cos \left( \frac{\Phi_i}{\Phi_0} \right)\\ \nonumber
&+& \frac{1}{2} C \dot\Phi_R^2 - \frac{\Phi_R^2}{2L} + \frac{1}{2} C_Q (\dot\Phi_1-\dot\Phi_2)^2 \\ \nonumber
&+& \sum_{i=1,2} \frac{1}{2} C_0 (\dot\Phi_i-\dot\Phi_R)^2.
\eeq
The associated Hamiltonian $H=H_0+H_1$ can be written as equations (\ref{H0}) and (\ref{H1}), with 
\beq
 \mathcal{C}&=&
\left(\begin{array}{ccc}
C_1 +C_Q + C_0  & - C_Q & - C_0 \\ \nonumber
-C_Q & C_2 +C_Q + C_0 & - C_0 \\
-C_0 & -C_0 & C+C_0
\end{array}\right), \\
 \mathcal{L}^{-1} &=&
\left(\begin{array}{ccc} \nonumber
1/L_1 & 0 & 0 \\
0 & 1/L_2 & 0 \\
0 & 0 & 1/L
\end{array}\right).
\eeq
Here, $L_i \equiv \frac{\Phi_0^2}{E_J^{(i)}}$. Quantizing this Hamiltonian gives
\beq
H &\approx& \sum_{i=1,2,R} \omega_i n_i -  \sum_{i=1,2} \Lambda_i n_i^2 + \\ \nonumber
&+& \sum_{i\neq j=1,2,R} \lambda_{ij} (a_i^{\dagger}a_j + {\rm h.c.}),
\eeq
with similar expressions for the coupling constants as in Eq.~(\ref{Hamilt}) \footnote{For the qubit-cavity coupling the capacitance $C_Q$ is substituted by $C_0$ in such expressions.}. Here, we will consider qubit and cavity far detuned. However, the qubit-cavity coupling may be sufficiently large to mediate virtual transitions between both qubits through the cavity. This effect will be relevant when $\frac{\lambda_{iR}\lambda_{jR}}{\Delta_j} \sim \lambda_{12}$, where $\Delta_j \equiv -\omega_j + \omega_R$. Performing a canonical transformation
\beq
H \to e^{S}He^{-S},
\eeq
with
\beq
S\equiv \sum_{i=1,2} \varepsilon_{iR} ( a_i^{\dagger}a_R - a_R^{\dagger}a_i),
\eeq
and taking $\varepsilon_{iR}\approx -\frac{\lambda_{iR}}{\Delta_i}$, we obtain, to second order in the transformation, an effective Hamiltonian (in a rotating frame with respect to the resonator)
\beq \label{Heff}
H_{\rm eff} &\approx& \sum_{i=1,2} \tilde\Delta_i n_i -  \sum_{i=1,2} \tilde\Lambda_i n_i^2 +\\ \nonumber
&+& \sum_{i\neq j=1,2} J_{ij} (a_i^{\dagger}a_j + {\rm h.c.}),
\eeq
with the renormalized constants are related to the original energies as
\beq
\tilde\Delta_i &=& -\Delta_i (1 - \varepsilon_{iR}^2), \\
\quad \tilde\Lambda_i &=& \Lambda_i (1+4\varepsilon_{iR}^2), \\
\quad J_{ij} &=& \lambda_{ij} + \frac{\lambda_{iR}\lambda_{jR}}{2\Delta_j}, 
\eeq
where $i=1,2$. The Hamiltonian (\ref{Heff}) can be readily generalized for $N$ qubits by letting $i=1,\ldots,N$, and conveniently diagonalized into the form written in Eq.~(\ref{diagH}). Also, it can be written in a spin language, by considering sufficiently large qubit anharmonicities. This allows us to replace the bosonic operators by spin 1/2 operators, so the effective Hamiltonian for two capacitively coupled superconducting Transmon qubits inside a cavity is given by
\beq
H_{\rm eff} \approx \sum_i \omega_i S_i^z + \sum_{i\neq j} J_{ij} ( S_i^{+}S_j^{-} + {\rm h.c.} ).
\eeq
Note that the angle and distance dependence of the qubit coupling is hidden in the $J_{ij}$ in this derivation.

\section{Qubit state measurement and Flux tuning}
An advantage of a circuit QED setup is the ability to measure the state of a predetermined but otherwise arbitrary subset of qubits in the lattice. A measurement of $\left<S^z_i\right>$ can be realised in circuit QED by probing the cavity transmission in the limit of a dispersive qubit cavity coupling. With the newly developed quantum limited amplifier technology it is nowadays possible to even detect quantum jumps of an individual qubit~\cite{SVijay2011,SHatridge2013}. Using this measurement procedure one has to take into account the strong interaction between the qubits, which means that a direct measurement of $\left<S^z_i\right>$ in the uncoupled basis is not possible~\cite{SGywat2006}. We would rather project onto one of the eigenstates of the coupled system which might be useful for an initial evaluation of the interaction but would not tell us anything about the actual properties of the state. 

This effect can be avoided by effectively switching of the coupling by detuning a subset of qubits from each other and all other qubits by more than $J$ in a time much faster than $1/J$ in order to be non adiabatic. This detuning can be achieved by realizing the qubits one wants to measure with a squid instead of a single Josepshon junction. The flux through these qubits can be adjusted with the help of small coils just outside the cavity. Simulations and preliminary experiments show that a current that is sufficiently large to generate the desired flux can be switched in less then $10~ns$. Changing the current in one of these coils will effect more than one qubit though with different magnitude. Adjusting the currents in the different coils accordingly using e.g. an optimal control algorithm should in principle allow us to generate any desired flux at the qubit location even though there are interdependencies.

\section{Scaling of the bond order parameter and bond correlations}

We provide here some additional information on the bond order parameter and correlation used in the main text to characterize the dimer phase (DP). The main message here is that, while the DP is rigorously characterized by the true order parameters $D^\alpha$, simpler correlation functions, which can immediately be accessed in the experiments, already qualitatively demonstrate the existence of the DP.

\begin{figure}[t]
%\begin{center}
\includegraphics[width = 0.65\columnwidth]{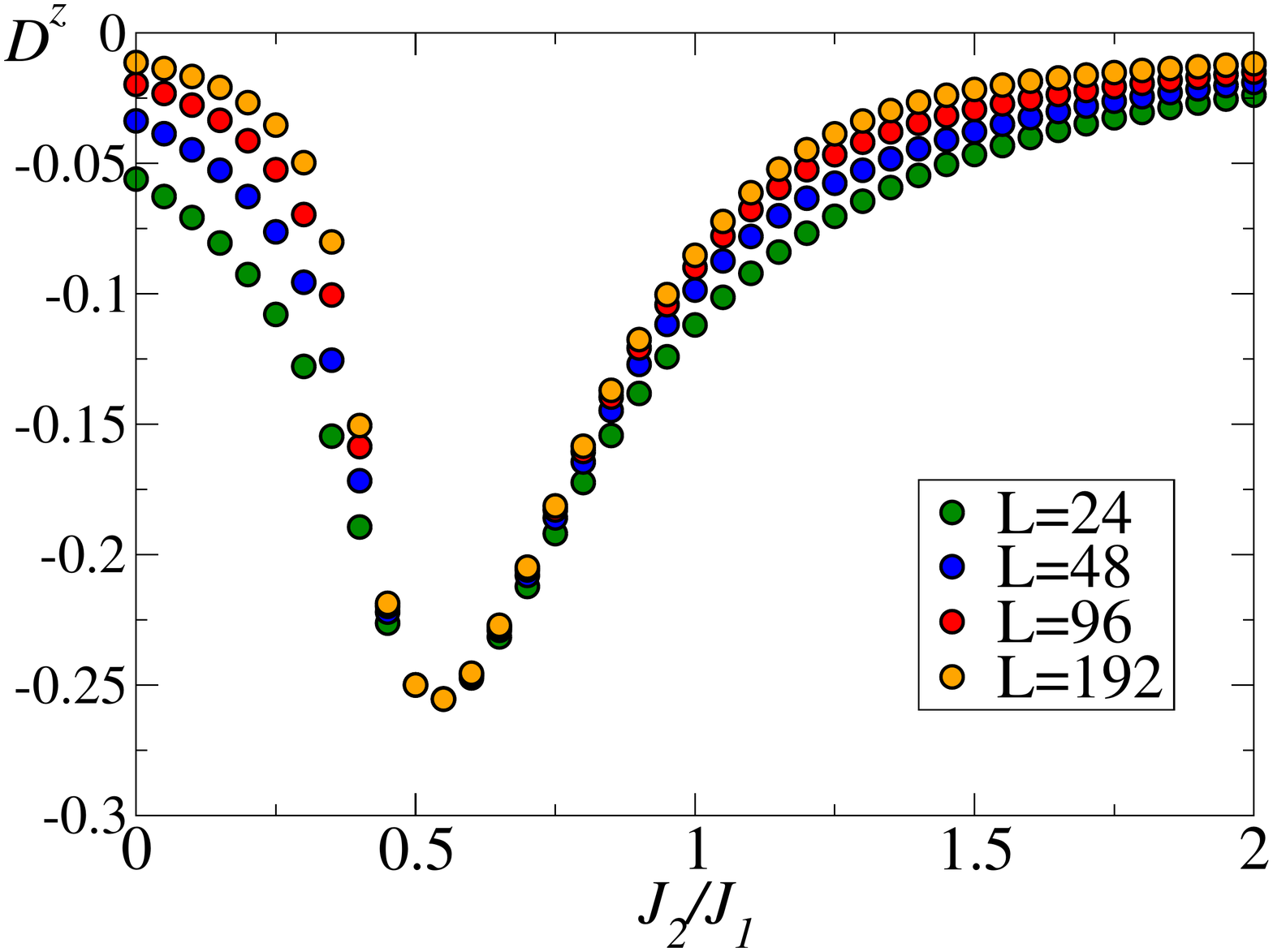}
\includegraphics[width = 0.65\columnwidth]{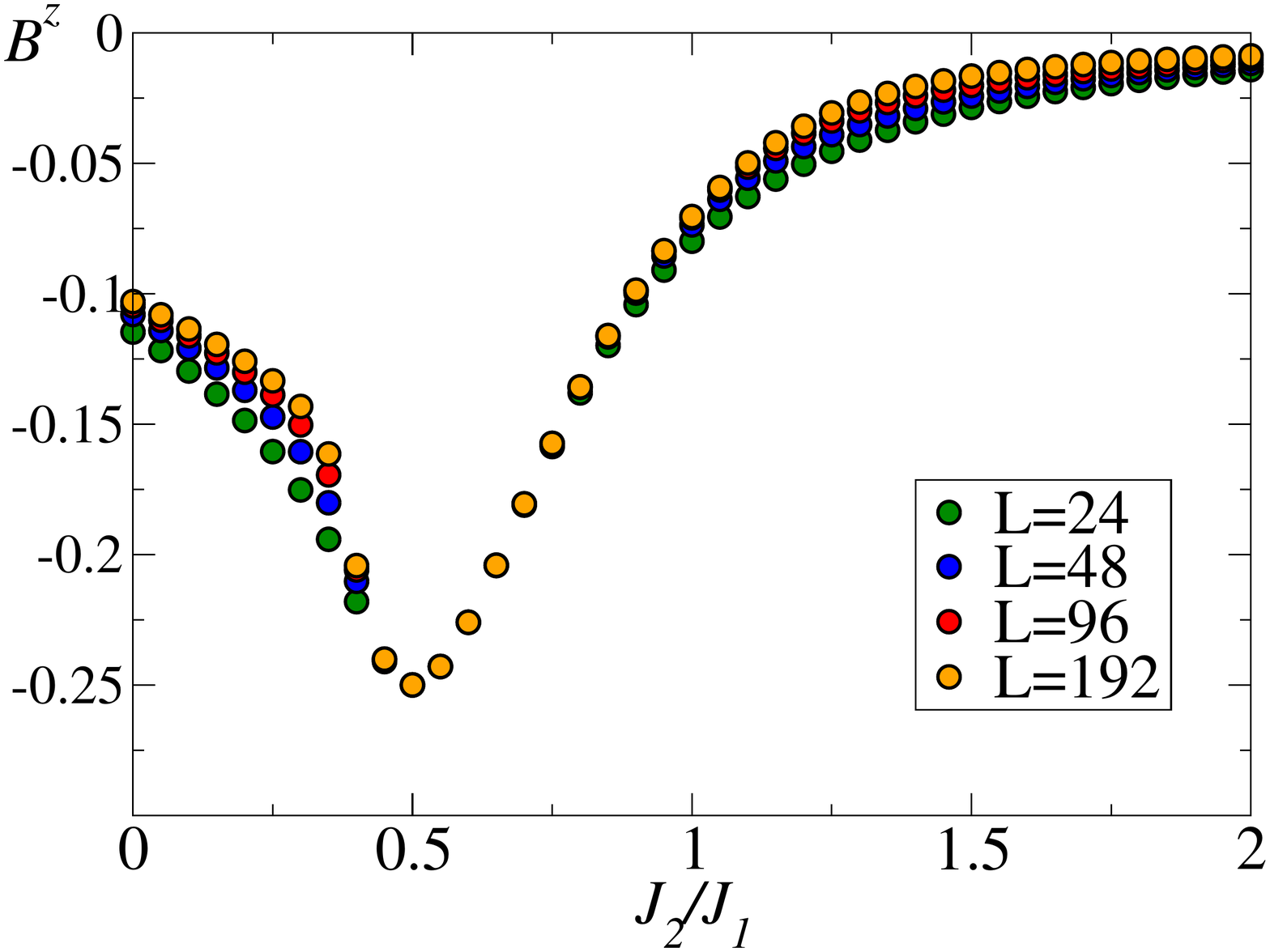}
\includegraphics[width = 0.65\columnwidth]{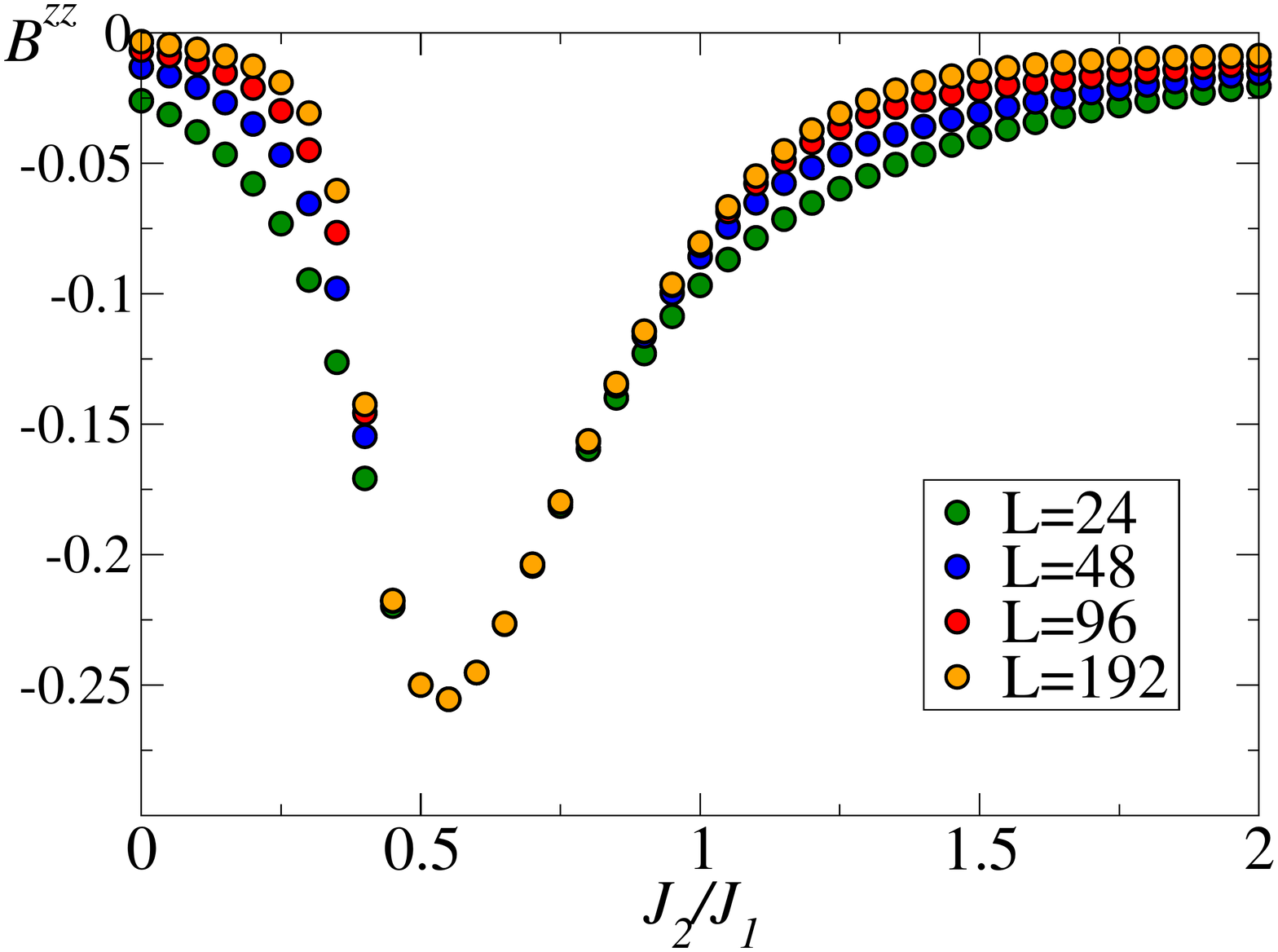}
\caption{From top to bottom: Scaling of $D^z, B^z, B^{zz}$ as a function of $J_2/J_1$ for different system sizes. }
\label{Bpars} 
\end{figure}

\begin{figure}[t]
%\begin{center}
\includegraphics[width = 0.75\columnwidth]{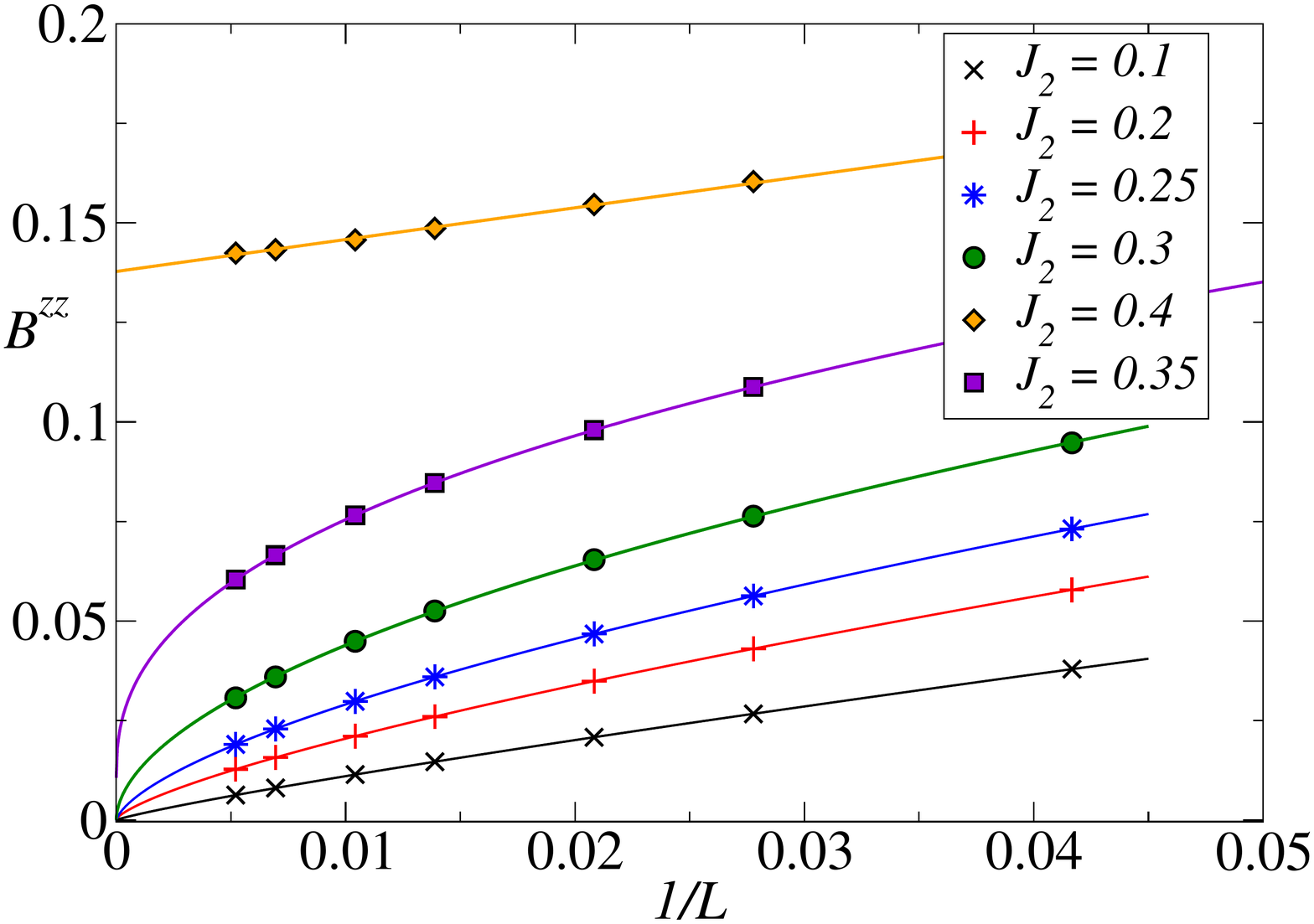}
\caption{Scaling of the $B_{zz}$ order parameter as a function of the inverse length of the system.}
\label{Bpars2} 
\end{figure}
As a first figure of merit, we show in top two panels of Fig.~\ref{Bpars} a comparison for different system sizes of the true bond order parameter $D^z$, and the bond correlation in the center of the chain $B^z$. As already stated in the text, these results further confirm that the bond correlation well captures the DP indicating that a dimer is indeed formed in the center bond of the chain. 

The local order parameter can be further simplified by taking the difference between the center bond an a neighbor, that is:
\begin{equation}
B^{zz} = D^z_{L/2} - D^z_{L/2+1}.
\end{equation}
This quantity has been extensively used as a true order parameter in numerical simulations for the bond order phase of extended Hubbard models (see, e.g., \cite{SEjima2007}), which have directly related to the DP here. In the lower panel of Fig.~\ref{Bpars}, we illustrate the scaling of $B^{zz}$ as a function of $L$ and different couplings. Despite its simplicity, this combination of correlation functions, which can be directly accessed in the setup discussed in the main text, can serve as an even better indicator of the DP with respect to $B^z$. This is further demonstrated by the results in Fig.~\ref{Bpars2}: there, we show how the finite-size scaling of $B^{zz}$ is also able to approximately detect the critical point between the DP and the superfluid phase in our model. An additional discussion on these types of order parameters can be found in Ref.~\cite{SKumar2010}

\bibliography{bibcQEDsupmat}

\end{document}